\documentclass[preprint]{ptephy_v1}
\preprintnumber{XXXX-XXXX}

\usepackage{graphicx}
\usepackage[utf8]{inputenc}
\usepackage[separate-uncertainty]{siunitx}
\usepackage{amsmath}
\usepackage{color}
\usepackage{ulem}
\usepackage{subcaption}
\usepackage{multirow}
\usepackage{hyperref}

\begin{document}

\title{In-situ high voltage generation with Cockcroft-Walton multiplier for xenon gas time projection chamber}

\author[1,*]{Shinichi~Akiyama}
\author[2]{Junya~Hikida}
\author[8]{Masashi~Yoshida}
\author[2]{Kazuhiro~Nakamura}
\author[3]{Sei~Ban}
\author[2]{Masanori~Hirose}
\author[1]{Atsuko~K.~Ichikawa}
\author[4]{Yoshihisa~Iwashita}
\author[2]{Tatsuya~Kikawa}
\author[6]{Yasuhiro~Nakajima}
\author[1]{Kiseki~D.~Nakamura}
\author[2]{Tsuyoshi~Nakaya}
\author[7]{Shuhei~Obara\thanks{Present Address: Institute for Advanced Synchrotron Light Source, National Institute for Quantum Science and Technology, Sendai 980-8579, Japan}}
\author[5]{Ken~Sakashita}
\author[8,9]{Hiroyuki~Sekiya}
\author[2]{Bungo~Sugashima}
\author[1]{Soki~Urano}
\author[1]{Sota~Hatsumi}
\author[1]{Sota~Kobayashi}
\author[2]{Hayato~Sasaki}

\affil[1]{Department of Physics, Graduate School of Science, Tohoku University, Sendai 980-8578, Japan \email{akiyama.shinichi.s8@dc.tohoku.ac.jp}}
\affil[2]{Department of Physics, Graduate School of Science, Kyoto University, Kyoto 606-8502, Japan}
\affil[3]{International Center for Elementary Particle Physics, University of Tokyo, Tokyo, 113-0033, Japan}
\affil[4]{Institute for Integrated Radiation and Nuclear Science, Kyoto University, Kumatori 590-0494, Japan}
\affil[5]{High Energy Accelerator Research Organization (KEK), Tsukuba 305-0801, Japan}
\affil[6]{Department of Physics, Graduate School of Science, University of Tokyo, Tokyo, 113-0033, Japan}
\affil[7]{Research Center for Neutrino Science, Frontier Research Institute for Interdisciplinary Sciences, Tohoku University, Sendai, 980-8578, Japan}
\affil[8]{Kamioka Observatory, Institute for Cosmic Ray Research, The University of Tokyo, Hida, 506-1205, Japan}
\affil[9]{Kavli Institute for the Physics and Mathematics of the Universe, The University of Tokyo, Kashiwa, 277-8583, Japan}

\begin{abstract}
We have newly developed a Cockcroft-Walton (CW) multiplier that can be used in a gas time projection chamber (TPC).
A TPC requires a high voltage to form an electric field that drifts ionization electrons.
Supplying the high voltage from outside the pressure vessel requires a dedicated high-voltage feedthrough.
An alternative approach is to generate the high voltage inside the pressure vessel with a relatively low voltage introduced from outside.
A CW multiplier can convert a low AC voltage input to a high DC voltage output, making it suitable for this purpose.

We have integrated a CW multiplier into the AXEL (A Xenon ElectroLuminescence detector), a high pressure xenon gas TPC to search for neutrinoless double beta decay of $^{136}$Xe.
It uses silicon photomultipliers to detect the ionization electrons through elecrtoluminescence, making it strong against electronic noise.
Operation of the CW multiplier was successfully demonstrated; the TPC was operated for 40 days at \SI{6.8}{\bar}, and an energy resolution as high as \SI{0.67+-0.08}{\%} (FWHM) at \SI{2615}{\keV} was obtained.
\end{abstract}
\subjectindex{H20}
\maketitle

\section{Introduction}
Neutrinoless double beta decay ($0\nu\beta\beta$) is a phenomenon in which two beta decays occur simultaneously in a same nucleus and the neutrinos annihilate each other virtually, resulting in the emission of only two electrons.
This phenomenon occurs when neutrinos have a Majorana nature, and, even if it happens, it would be a very rare decay\cite{PhysRev.56.1184}.
So far, it has never been observed.
One of the candidate parent nuclides is $^{136}$Xe, for which the KamLAND-ZEN experiment has given the most stringent lower limit of the half-life: \num{2.3e26}~years (90\% C.L.) for this particular type of decay\cite{PhysRevLett.130.051801}.
The corresponding upper limit on the effective Majorana neutrino mass is in the range 36-156~meV.
In order to conduct experiments with higher sensitivity, a ton-scale detector with better energy resolution and lower background is required.
High-pressure xenon gas time projection chambers (TPCs) are detectors that fulfill these requirements.
However, there are also technical challenges.
One such challenge is related to the high voltage to generate the electric field to guide ionization electrons to the detection surface.
For a \SIrange{100}{1000}{kg} gas TPC, the required voltage would be higher than \SI{100}{\kV}.
To feed such a high voltage (HV) from outside the pressure vessel, high voltage feedthroughs compatible with high pressures are needed.
Another approach is to introduce a relatively low voltage from outside the pressure vessel and boost it inside the pressure vessel.
The Cockcroft-Walton (CW) multiplier\cite{doi:10.1098/rspa.1932.0107} can be used to convert a low voltage AC input to a high voltage DC output.
This approach was proposed for liquid argon TPC's\cite{Marchionni_2011}, but has not been realized in actual operation.
One difficulty comes from the large baseline shift caused by the AC input, which makes signal readout and analysis difficult.
If the AC input is turned off after charging the capacitors in the CW multiplier, this baseline shift disappears but it has to be shown that the voltage is properly maintained.
 
We are developing a high pressure xenon gas TPC called AXEL, A Xenon ElectroLuminescence detector.
We use the electroluminescence (EL) process as a signal: electrons accelerated by an electric field excite atoms, and photons are emitted as a result of the subsequent de-excitation.
The idea of using the EL process in a high-pressure gas TPC's to achieve high energy resolution and tracking performance was first proposed and adopted by the NEXT experiment\cite{4437191}\cite{VÁlvarez_2013}, and $0\nu\beta\beta$ search has already been conducted with this technology by them\cite{Novella2023}.
Since the ionization signal is converted to light to be read out and the light signal is amplified with quite high efficiency by photon counters, it is highly resistant to electronic noise.
We have developed a CW multiplier to supply high voltage to the AXEL detector\cite{10.1093/ptep/ptad146} and installed it at the \SI{180}{\L} prototype detector.
The performance was evaluated at the \SI{2615}{\keV} $\gamma$-ray peak from $^{208}\mathrm{Tl}$, through long-term measurements. 
This paper reports on the developed CW multiplier, and the performance of the TPC equipped with it.

\section{AXEL detector}
\begin{figure}[tb]
  \centering
  \includegraphics[width=0.95\linewidth]{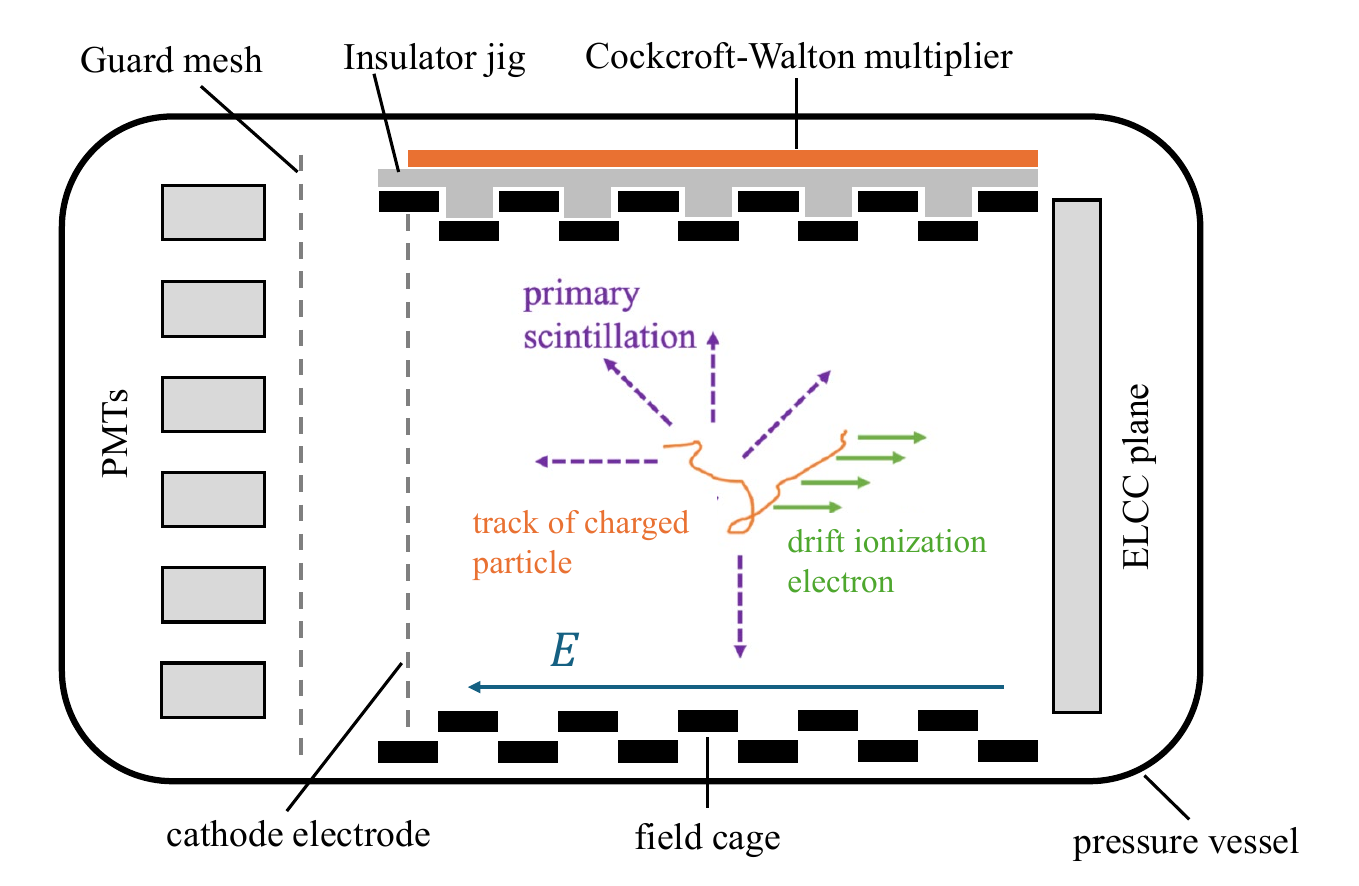}
  \caption{Schematic view of the AXEL \SI{180}{\L} prototype detector.}
  \label{fig:detector_overview}
\end{figure}

A schematic view of the AXEL \SI{180}{\L} prototype detector is shown in Fig.~\ref{fig:detector_overview}.
As a charged particle passes through the gas volume of the detector, it excites and ionizes xenon atoms along the track.
The excited atoms result in emissions of the primary scintillation light, which is detected with photomultiplier tubes (PMTs).
This signal is used as the timing of the event and recorded with a 14-bit 100 MS/s digitizer (CAEN, v1724).
A guard mesh connected to the ground is placed in front of the PMTs to protect the PMTs from the cathode voltage.
Ionization electrons drift along the uniform drift electric field formed by a field cage toward the detection plane called electroluminescence light collection cell (ELCC) plane.
The ELCC is a cellular ionization-electron detection device utilizing the EL process described in Sec.~\ref{sec:elcc}.
The CW multiplier is mounted on the insulator jig to supply high voltage to the field cage.
The details of the CW multiplier and the field cage are described in Sec.~\ref{sec:CW} and Sec.~\ref{sec:field_cage} respectively.
The time for ionization electrons to drift through the chamber is of the order of \SI{10}{\us}, whereas the scintillation light is sufficiently short $\sim$\SI{100}{\ns}.
This readout in each cell of the ELCC plane with time information allows reconstruction of the 3D track.
Since the stopping power of the xenon gas for electrons is proportional to the density of the xenon gas, the length of the track is typically inversely proportional to the density of the xenon gas.
The number of detected photons at each timing and cell position gives the energy deposit of the corresponding position on the track.
The total number of photons gives the total energy deposited along the track.
The detailed configuration of the AXEL detector is described in a previous work\cite{10.1093/ptep/ptad146}.
In the following, we describe the ELCC and the field cage in detail, which were improved from \cite{10.1093/ptep/ptad146}.

    \subsection{ELCC}\label{sec:elcc}
\begin{figure}[tb]
        \centering
        \includegraphics[width=0.7\linewidth]{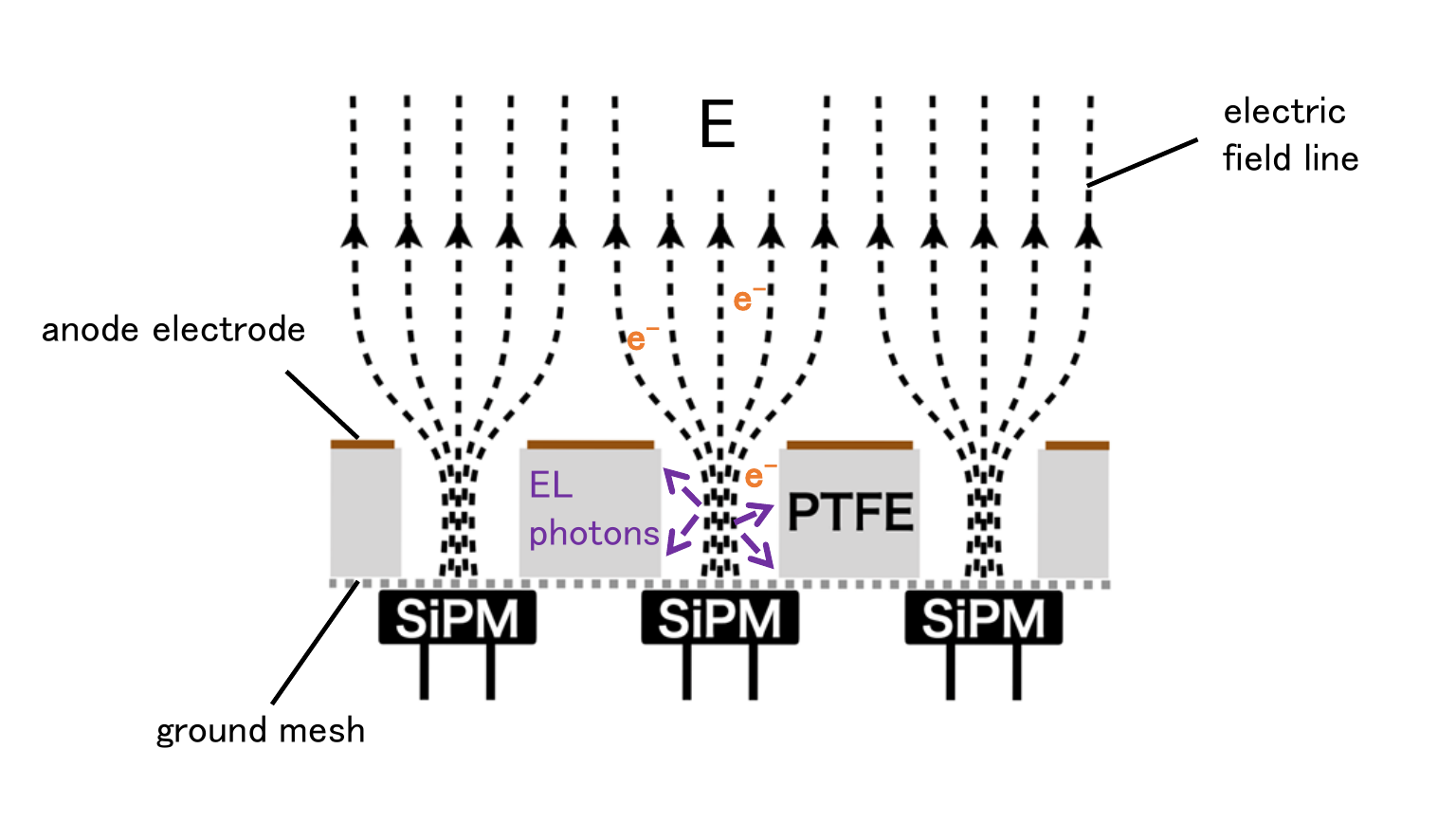}
        \caption{Schematic cross-sectional view of the ELCC.}
        \label{fig:elcc_schematic}
    \end{figure}

\begin{figure}[tb]
        \centering
        \includegraphics[width=0.6\linewidth]{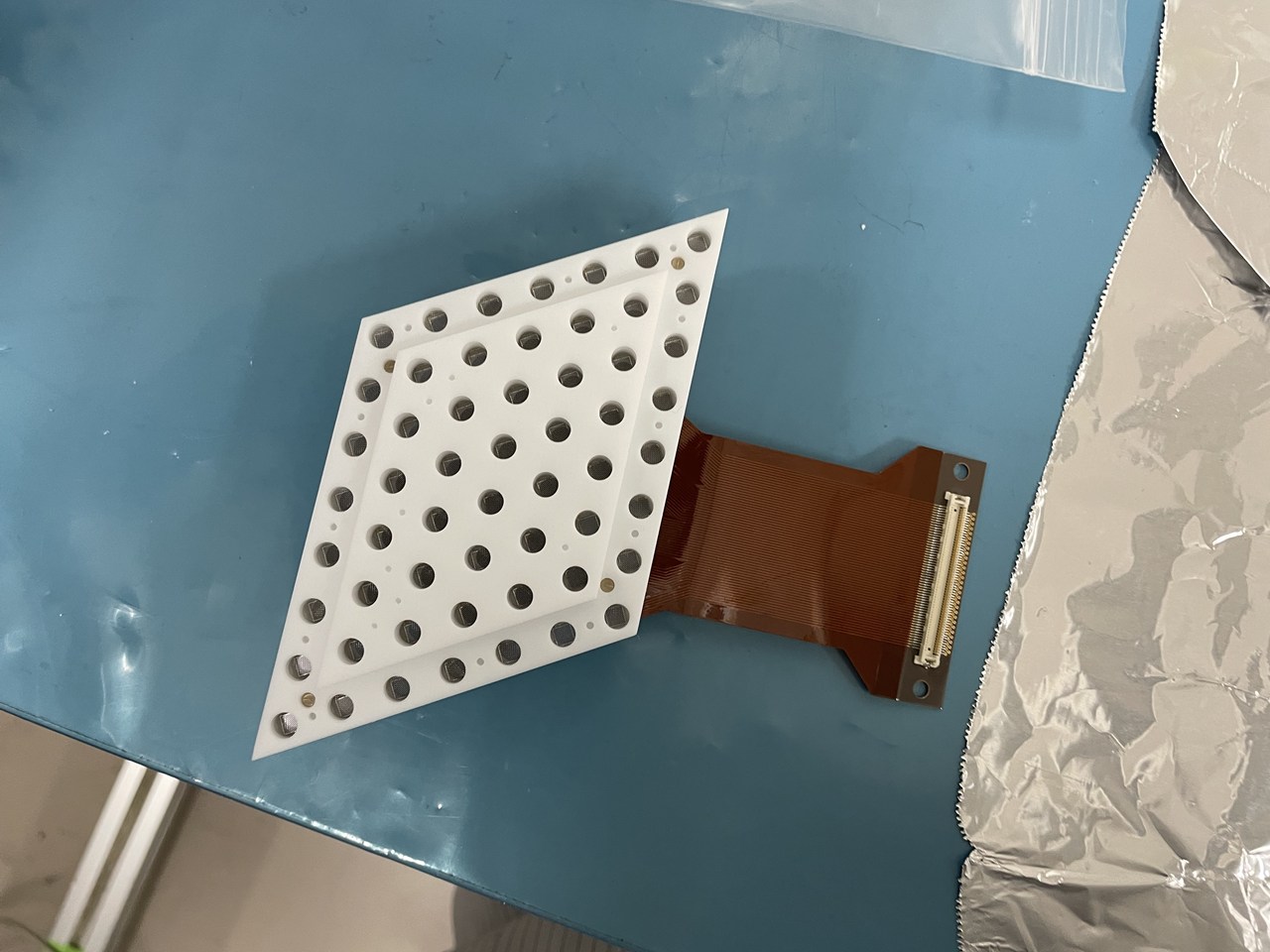}
        \caption{A picture of an ELCC unit without the anode electrode. One cell of the perimeter is indented for the cover layer. One unit consists of $56~(=7\times8)$ cells. This unit is laid out to form the ELCC plane.}
        \label{fig:elcc_unit}
    \end{figure}

    A schematic cross-sectional view of the ELCC is shown in Fig.~\ref{fig:elcc_schematic}.
    The ELCC consists of the TPC anode and a ground mesh electrode, with a 5mm-thick polytetrafluoroetylene (PTFE) plate between them.
    The plate has round holes arranged in a hexagonal pattern at \SI{1}{\cm} intervals, as the cells.
    Ionization electrons, following the electric field lines, enter the cells in the ELCC.
    The inside of the cell is subjected to an electric field of \SI{3}{\kV/\cm/\bar} by the voltage difference between the anode and the ground mesh, where the field is strong enough to induce EL from the entering ionization electrons.
    The vacuum ultraviolet (VUV) photons emitted by the EL process in a cell are detected by the cell's SiPM, a Hamamatsu VUV sensitive S13370-3050CN multi-pixel photon counters (MPPC), located behind the ground mesh.
    The signals of MPPCs are stored at a sampling interval of 5 MS/s by a 12-bit ADC on a front-end board AxFEB\cite{9072179} connected through a flexible printed cable.
    The ELCC plane consists of multiple ELCC units, each composed of $56~(=7\times8)$ cells (Fig.~\ref{fig:elcc_unit}).
    For the \SI{180}{\L} prototype detector, 12 ELCC units are used.
    
\begin{figure}[tb]
        \begin{minipage}{\linewidth}
            \centering
            \includegraphics[width=\linewidth]{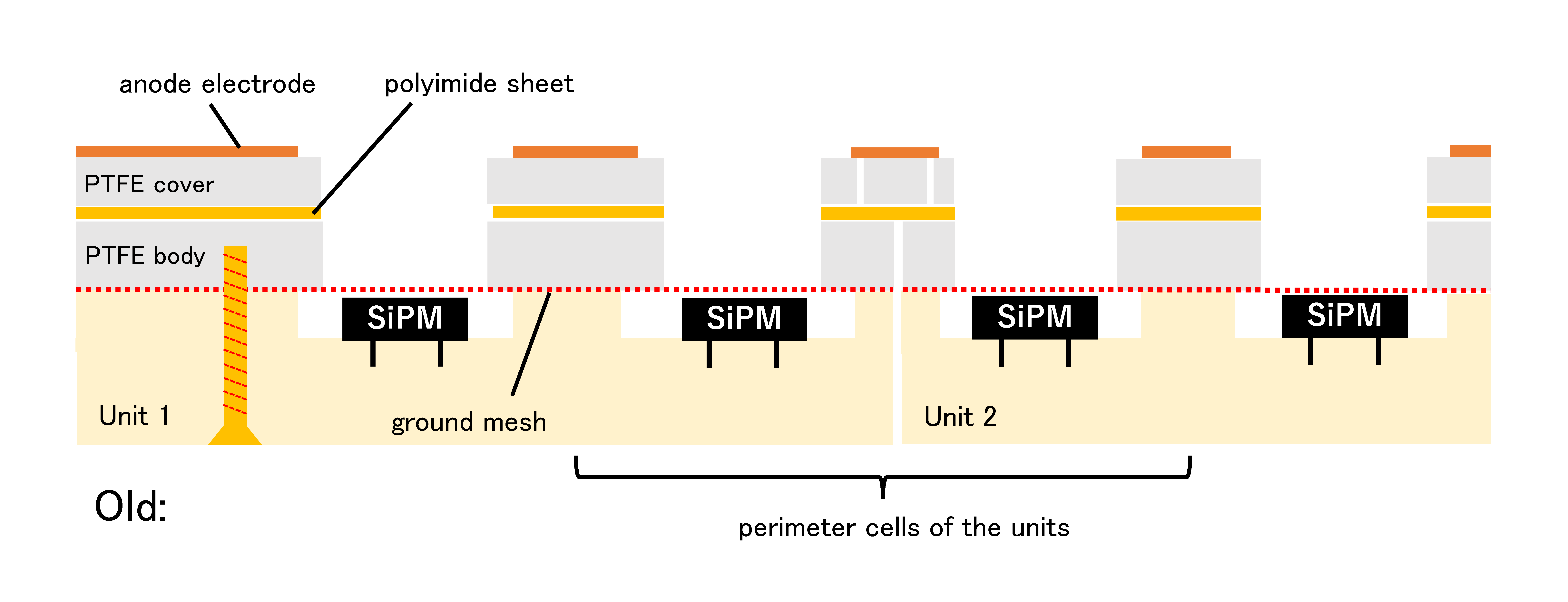}
        \end{minipage} \\
        \begin{minipage}{\linewidth}
            \centering
            \includegraphics[width=\linewidth]{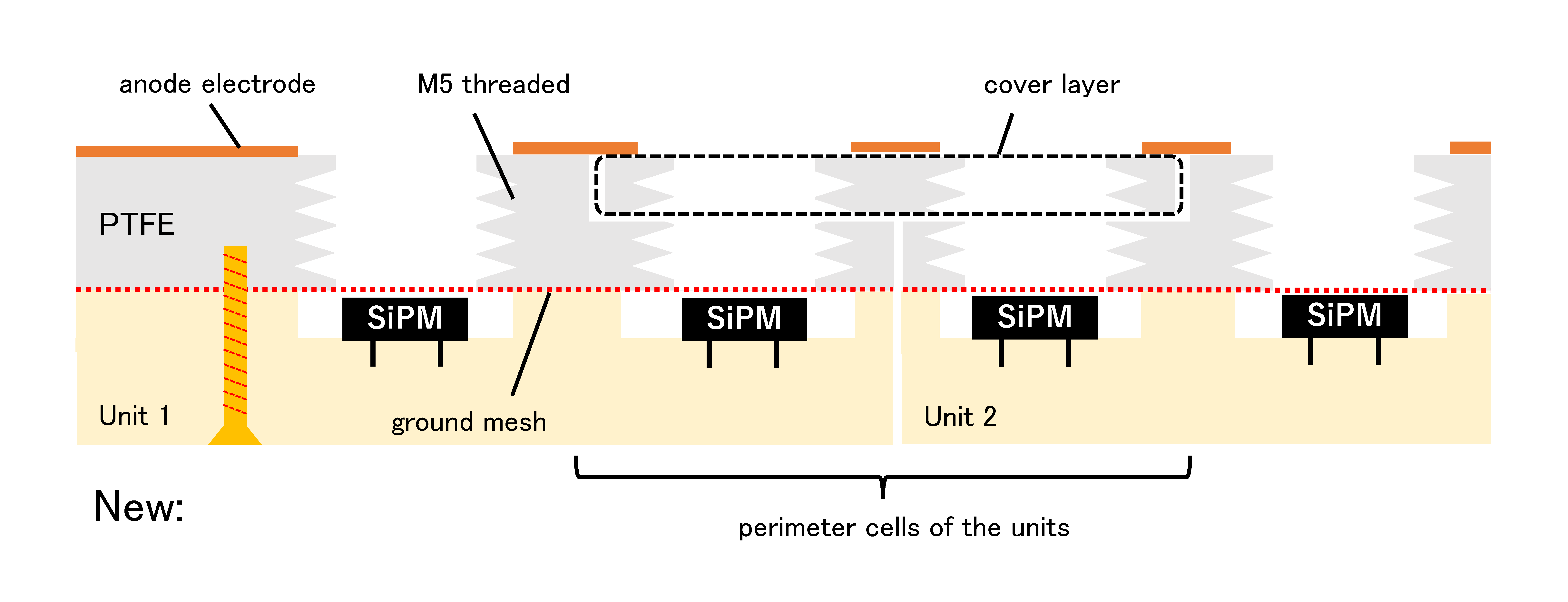}
        \end{minipage}          
        \caption{Schematic cross-sectional view of the ELCC structures. The two-layer structure of the previous study\cite{10.1093/ptep/ptad146} is shown on the top, and the tapped hole structure of the amplification section, the current upgrade, is shown on the bottom.}
        \label{fig:elcc_crosssectional}
    \end{figure}

    The target value of the field strength in the cell is \SI{3}{\kV/\cm/\bar} but has not been achieved due to discharges between the anode and the ground electrodes.
    In the previous study\cite{11.1093/ptep/ptaa030}, discharge occurred between the anode and the ground mesh electrodes at the boundaries of the ELCC units and at the screw holes to fix the ELCC units.
    As a countermeasure against these discharges, the PTFE plate was divided into two layers; a cover layer was added to cover the gaps between units.
    Also, the screws fixing the ELCCs were modified so that they do not penetrate the PTFE plate.
    Figure \ref{fig:elcc_crosssectional} top shows these modifications.
    However, electrical discharges had still occurred, so additional countermeasures are applied in this study.
    The two-layer structure in the previous study showed discharges presumably on the surface of the polyimide sheet sandwiched between the layers.
    Therefore, polyimide sheets are not used, and the two-layer structure is employed only at the perimeter cells of the unit.
    In addition, to prevent discharge along the holes of the cell, the holes are tapped using JIS M5 thread as shown in Fig.~\ref{fig:elcc_crosssectional} bottom.
   
    \subsection{Field cage}\label{sec:field_cage}
    The field cage is a frame for creating a uniform electric field that guides ionization electrons to the detection plane.
    The field cage of the \SI{180}{\L} prototype consists of two sizes of D-shaped band electrodes with diameters of about \SI{500}{\mm} arranged alternately with a \SI{10}{\mm} pitch.
    The flat section of the electrodes allows the CW multiplier to be placed between the chamber and the electrode.
    A \SI{15}{\mm} thick high-density polyethylene (HDPE) tube is inserted between the field cage and the chamber for insulation.
    The topmost electrode is the cathode with a mesh to allow scintillation light to pass through while preventing leakage of the electric field.
    The number of electrodes increased from 18 to 40 in order to expand the sensitive volume.
    The first 20 bands are made of \SI{3}{\mm} thick aluminum, and the rest is made of \SI{1.5}{\mm} thick copper.
    Aluminum was adopted for a cost reason.
    The voltage to the cathode electrode and each stage of the field cage was supplied by the CW multiplier described in Sec.~\ref{sec:CW}.

\section{Design and performance of the Cockcroft-Walton multiplier}\label{sec:CW}

    \subsection{Size and outgassing constraints}
\begin{figure}[tb]
        \centering
        \includegraphics[width=0.9\linewidth]{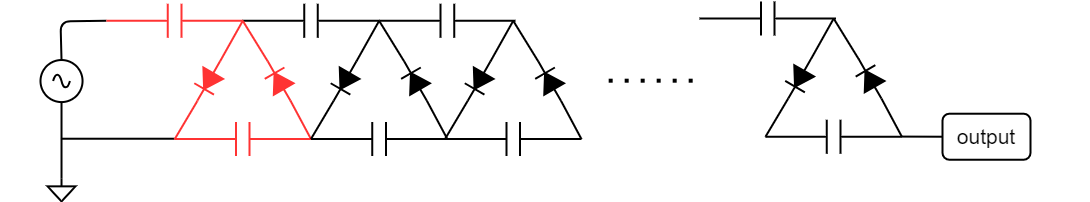}
        \caption{Schematic diagram of the CW multiplier.}
        \label{fig:cw_schematic}
    \end{figure}

    The schematic diagram of the CW multiplier is shown in Fig.~\ref{fig:cw_schematic}.
    The CW multiplier is composed of multiple stages of a basic circuit, consisting of two capacitors and two diodes, connected in series.
    The ideal output voltage $V$ of an $N$-stage CW multiplier is given as $V = 2NU$, where $U$ is the amplitude of the AC input voltage, but the actual output voltage of a CW multiplier is lower than the ideal voltage due to the parasitic capacitance of diodes and load resistance\cite{10.1063/1.1683930}.
    
    The AXEL experiment aims to apply a drift electric field of \SI{-800}{\V/\cm} on top of the \SI{-12}{\kV} across the ELCC in xenon at 8 bar.
    To install a CW multiplier inside the chamber, there are several requirements regarding size and material.
    The CW multiplier is installed in a narrow space between the HDPE tube and the field cage.
    In case of the \SI{180}{\L} prototype, the width is limited to about \SI{20}{\cm} and the height to \SI{3}{\cm}.
    The length is also limited to \SI{40}{\cm} to fit to the length of the field cage, and the target voltage is \SI{-44.8}{\kV}.
    
    The CW has to be composed of low outgassing materials.
    This is because impurities with high electronegativity, such as oxygen, adsorb ionization electrons, reducing the TPC signal and degrading the energy resolution. 
    The typical outgassing rate of the prototype detector was \mbox{\SI{1.23e-4}{\Pa.\m^3/\s}}\cite{10.1093/ptep/ptad146}, with which xenon gas purity was kept sufficiently high by continuous purification using a molecular sieve and a getter.
    The outgassing of the CW multiplier should be well below this rate.
    
    \subsection{CW components}
\begin{figure}[tb]
        \centering
        \includegraphics[width=0.6\linewidth]{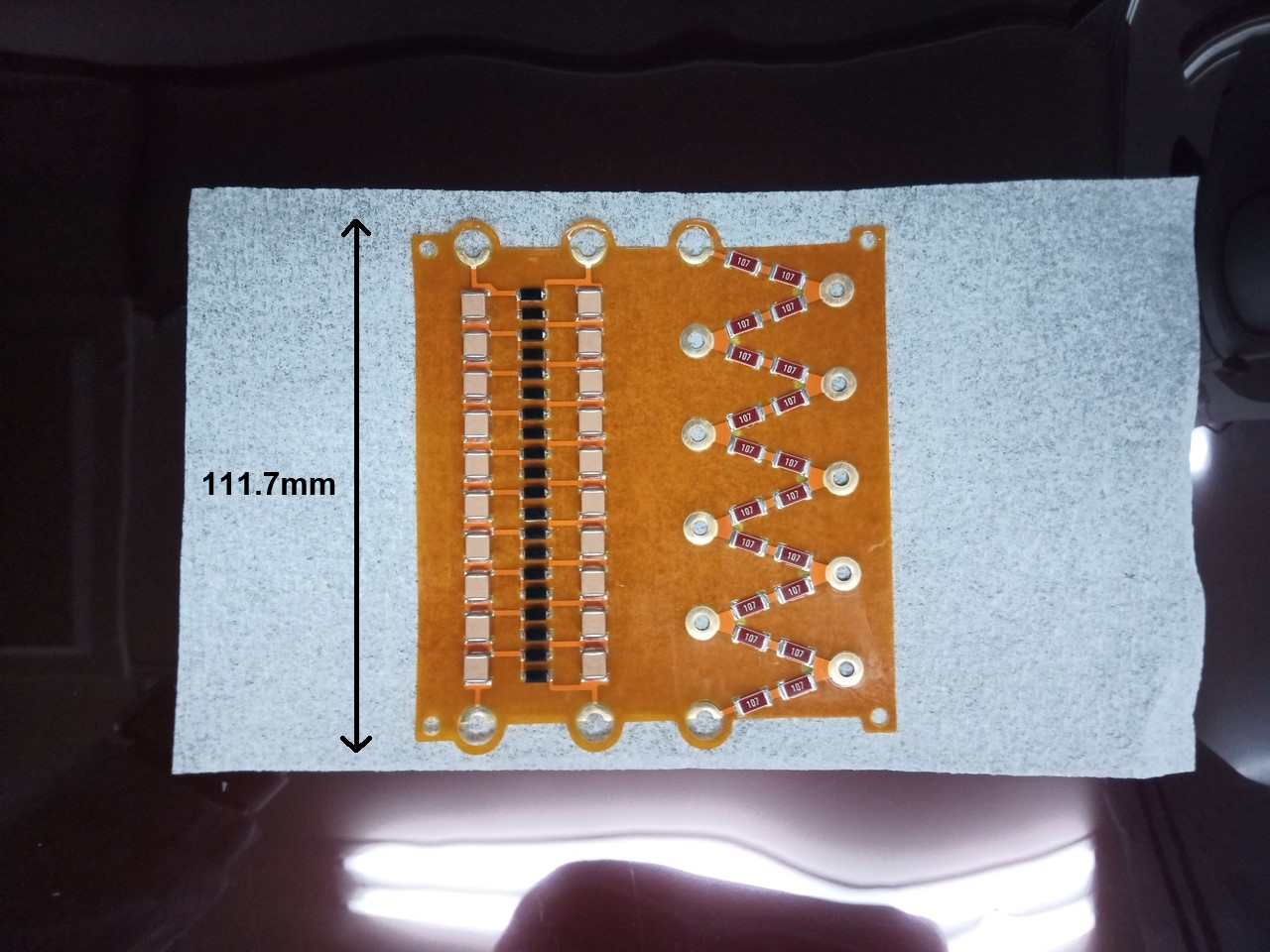}
        \caption{CW multiplier and resistor chain board implemented as a flexible printed circuit.}
        \label{fig:cw_sheet}
    \end{figure}

    To realize small dimensions and reduce outgassing, we have adopted a flexible printed circuit (FPC) board with surface-mount devices on it.   
    The FPC consists of an \SI{18}{\um}-thick copper electrode sandwiched between a \SI{25}{\um}-thick polyimide base and coverlay, NIKKAN INDUSTRIES F-30V and CISV respectively.
    One board is \SI{111.7}{\mm} long and contains 10 CW stages, and a resistor chain to evenly divide the electrical potential to the electrodes of the field cage.
    The FPC board has C-shaped terminals at both ends, allowing them to be connected to each other (Fig.~\ref{fig:cw_sheet}).
    This enables the construction of a CW circuit with more stages.
    The fewer the number of CW stages, the more efficient the voltage boosts, but the higher the voltage applied to the element per stage.
    Since the voltage rating of the capacitors, diodes and resistors is \SI{2}{\kV}, the circuit was set up with 10 stages per board.
    When the full, rated voltage is applied to these boards, the required total voltage of \SI{-44.8}{\kV} can be produced with an efficiency of about \SI{60}{\%}.
        
    We used a Knowles Syfer's \SI{0.1}{\uF} 2220 chip size ceramic capacitor.
    For the diodes we used fast switching diode Micro Commercial Components FM2000GP with a reverse recovery time of \SI{500}{\ns} and typical junction capacitance of \SI{8}{\pF}.
    For the resistor chain Bourns Inc. CHV2512-JW-107ELF \SI{100}{\mega\ohm} were used.

    \subsection{High voltage generation}
\begin{figure}[tb]
        \begin{minipage}{0.4\linewidth}
            \centering
            \includegraphics[width=\linewidth]{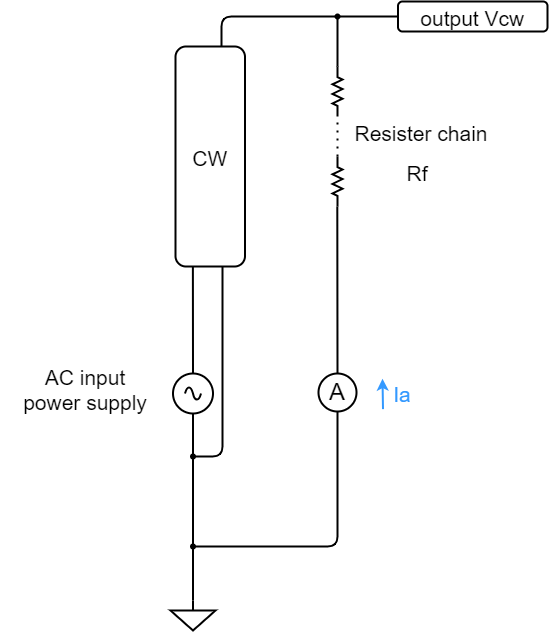}
        \end{minipage}
        \begin{minipage}{0.6\linewidth}
            \centering
            \includegraphics[width=\linewidth]{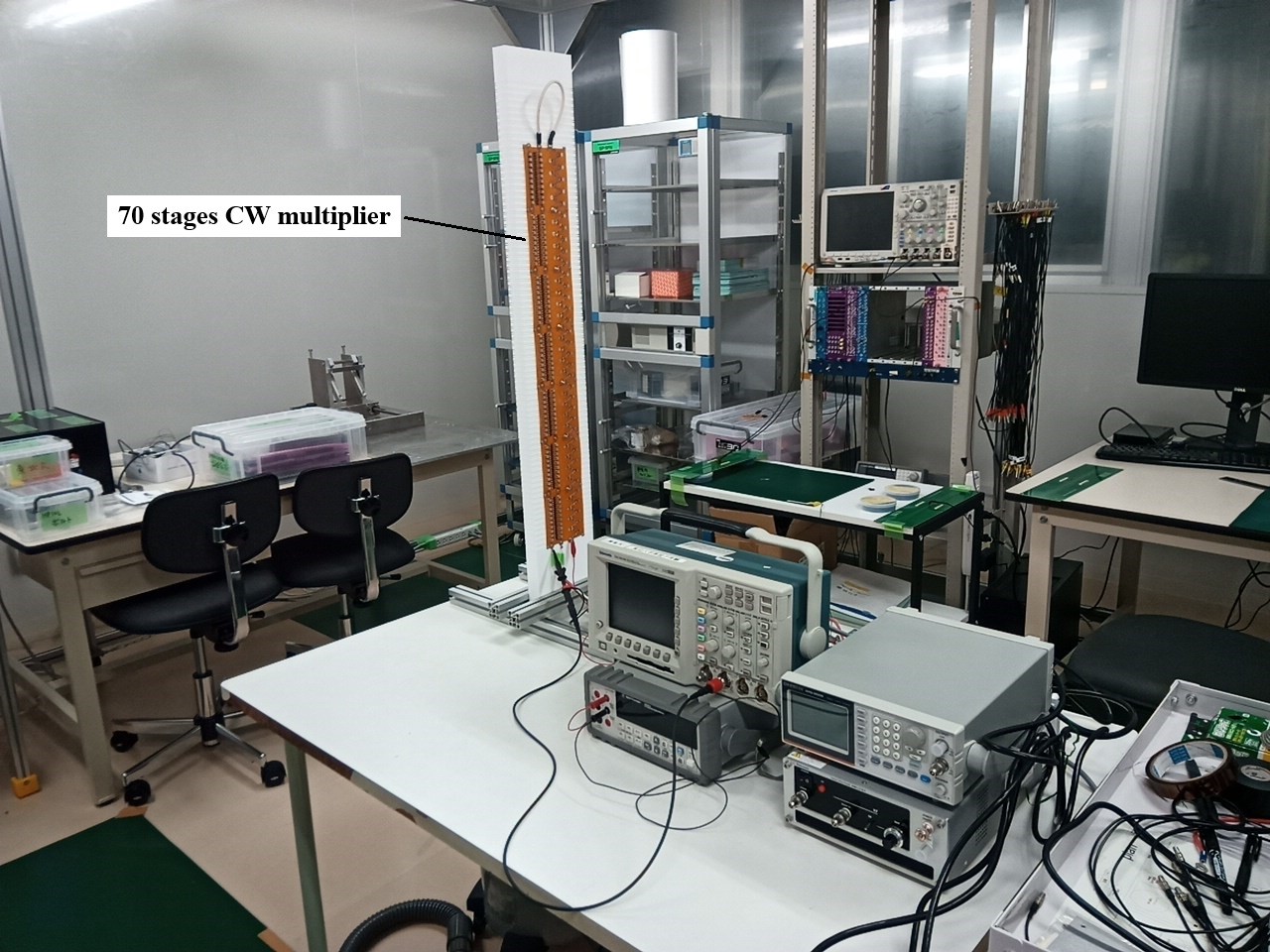}
        \end{minipage}          
        \caption{Schematic diagram (left) and a picture (right) of the measurement of the CW multiplier high voltage generation.}
        \label{fig:cw_readout_desktop}
    \end{figure}

    The voltage drop due to the load resistance is suppressed with higher input frequency.
    We measured the frequency dependence of the output voltage with the setup shown in Fig.~\ref{fig:cw_readout_desktop}.
    To feed the AC input power to the CW multiplier, a sine wave from a function generator GW Instek AFG-2005v was amplified by an AC amplifier Matsusada HJOPS-2B10.
    The maximum output voltage and current of the AC amplifier is \SI{+-2}{\kV}, \SI{10}{\mA} respectively.
    The slew rate is \SI{150}{\V/\us} and bandwidth is \SI{18}{\kHz} (\SI{-3}{dB}).
    The output of the CW multiplier was connected to the ground through a resistor chain $R_f$ on the boards and an ampere meter Agilent Technologies U3401A.
    The output voltage $V_{CW}$ is obtained from the measured current value $I_a$ as $V_{CW} = - I_a \times R_f$.
    Here the resistance value $R_f$ is \SI{2}{\giga\ohm} per 10 steps of the resistor chain on the boards.
    We prepared seven FPC boards for measurement.
    Before making the frequency measurement, a test was performed with an input peak-to-peak voltage of \SI{1800}{\V}, near the voltage rating, to the CW multiplier in a same setup that in Fig.~\ref{fig:cw_readout_desktop}.
    Because discharges happened on two boards during these tests, frequency measurements were only made on the remaining five boards.

\begin{figure}[tb]
        \begin{minipage}{0.5\linewidth}
            \centering
            \includegraphics[width=\linewidth]{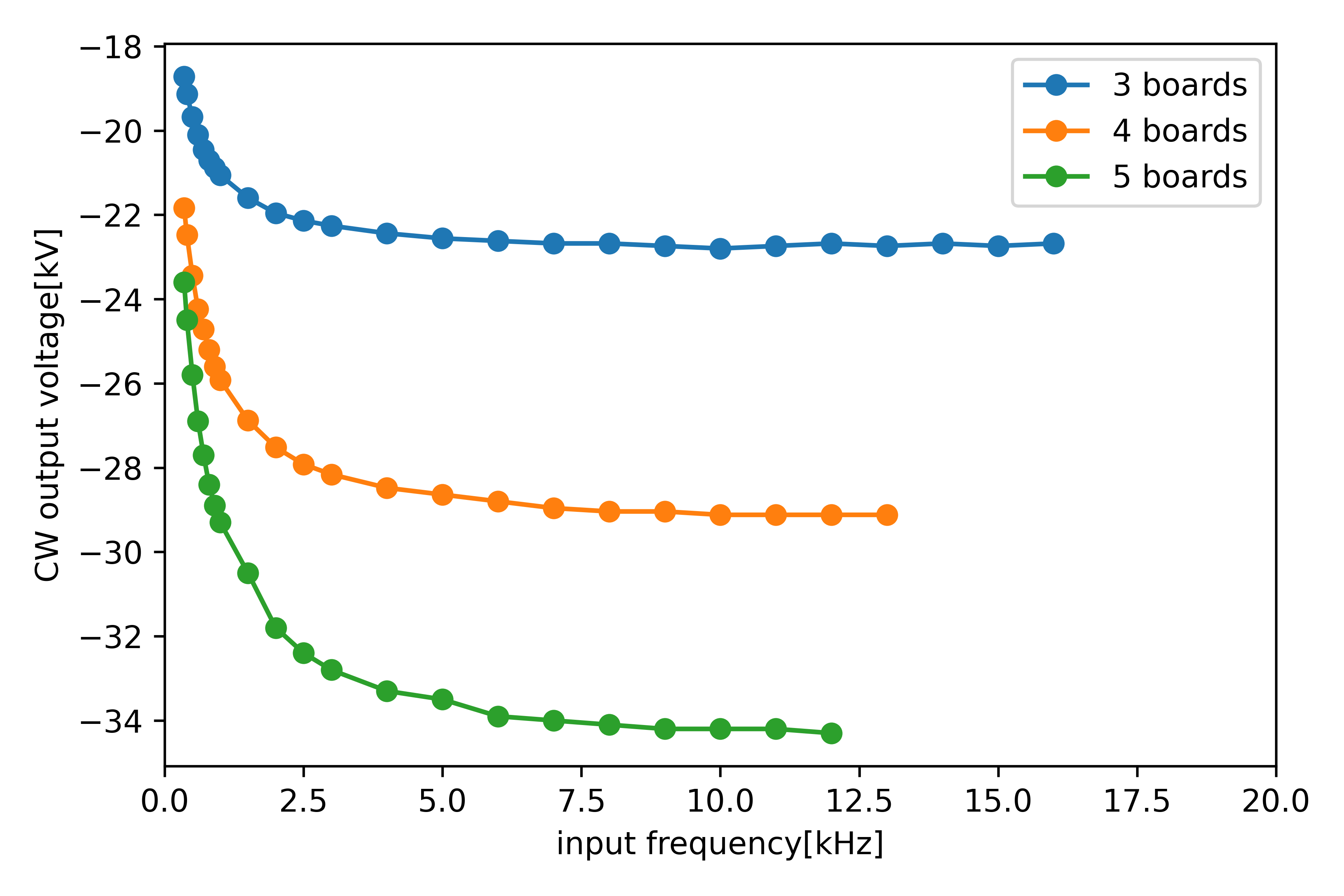}
        \end{minipage}
        \begin{minipage}{0.5\linewidth}
            \centering
            \includegraphics[width=\linewidth]{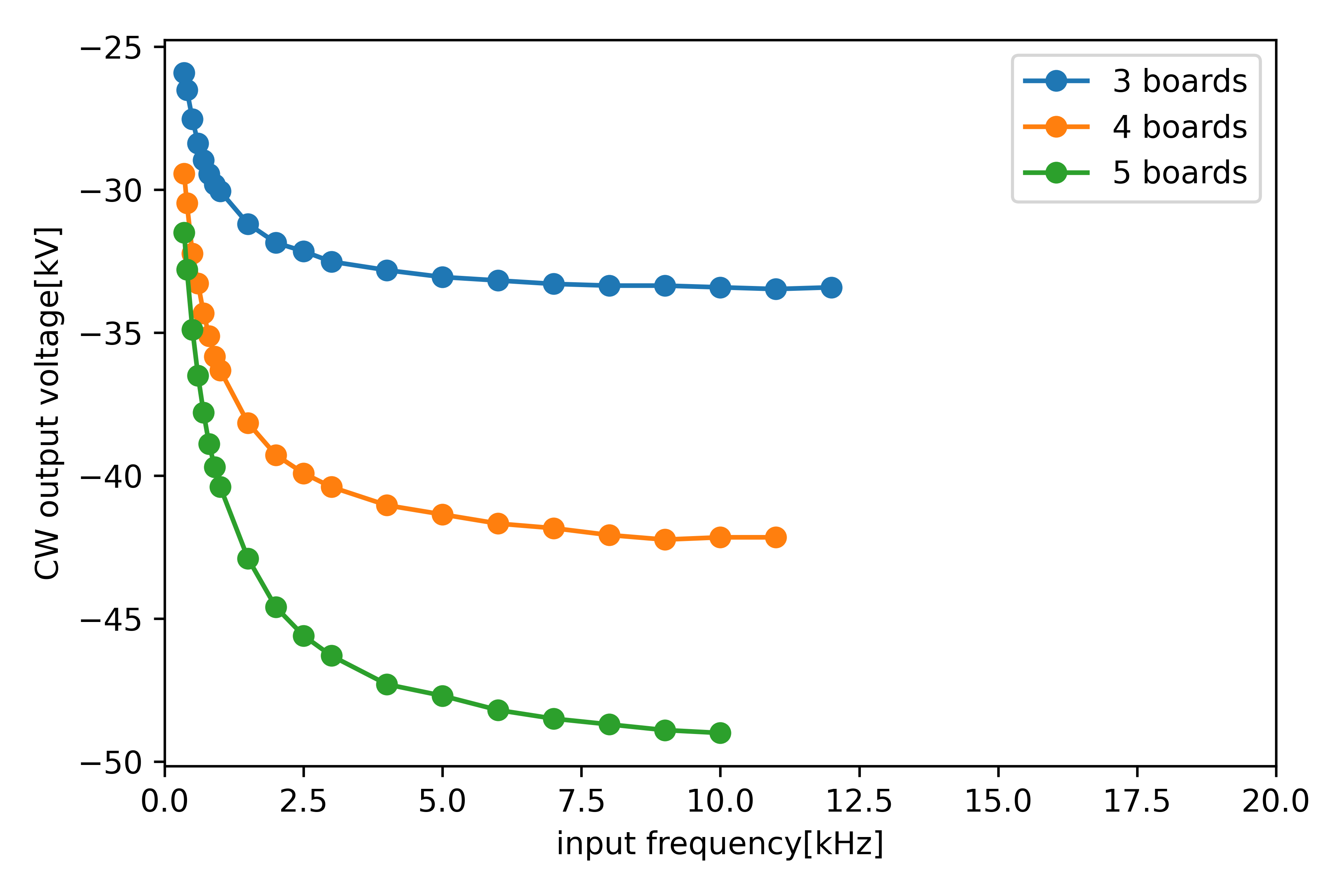}
        \end{minipage}          
        \caption{CW output voltages as a function of input frequency for \SI{800}{\V} peak-to-peak input (left) and for \SI{1200}{\V} (right).}
        \label{fig:cw_output_desktop}
    \end{figure}

    The measurement results are shown in Fig.~\ref{fig:cw_output_desktop}.
    The input peak-to-peak voltages are \SI{800}{\V} and \SI{1200}{\V}.
    Data points missing at high frequencies are due to instabilities in the amplifier output, which caused oscillations.
    As more boards were added, the frequency at which that happened became progressively lower.
    This is thought to be due to the increase in capacitive load with more boards, which leads to the amplifier's output current reaching its limit.
    We have confirmed that the operation at higher frequencies is possible if higher power audio amplifiers and transformers are used, and we plan to introduce this in the future.

    \subsection{CW surface insulation}
    During the test operation of the CW multiplier at a peak-to-peak voltage of \SI{1800}{\V} a surface discharge occurred on the capacitor and FPC surface and the output voltage of the CW multiplier deteriorated.
    To allow for higher voltage in- and output, the circuit was coated with the methyl silicone resin, Shin-Etsu Chemical Co. Ltd. KR-251, which was applied as follows:
    \begin{enumerate}
        \item Ultrasonically clean the circuit in ethanol for 15 minutes.
        \item Submerge the circuit in a vat containing KR-251 and run it through ultrasonic cleaning for 15 minutes to defoam.
        \item Degas the circuit under vacuum for one hour while submerged in KR-251.
        \item Lift the circuit out of the KR-251 bath and dry it under vacuum for 4 hours.
    \end{enumerate}
    Even with the above procedure, air bubbles around the circuit elements could not be completely avoided.
    To estimate the impact of outgassing from these bubbles on gas purity, we measured the outgassing rate of a coated FPC.
    
    \subsection{Outgassing rate of the coated CW circuit board}
    A vacuum test was performed using an FPC board that was cut into about one-fourth and then coated.  
    The cut board was placed in a NW50 pipe connected to a turbo molecular pump, evacuated for about 5 days.
    The outgassing rate was estimated from the pressure rise as a function of time after closing off the vacuum connection to the pump.
    The pressure changed from \SI{1.7e-4}{\Pa} to \SI{91}{\Pa} over about 14 hours.
    The estimated outgassing rate is \SI{8.9e-7}{\Pa.\m^3/\s} including leaks in the vacuum system.
    Since the \SI{180}{\L} prototype detector uses four circuit sheets, the outgassing is estimated to be \SI{1.4e-5}{\Pa.\m^3/\s}, which is about an order of magnitude less than the operational outgassing rate \mbox{\SI{1.23e-4}{\Pa.\m^3/\s}} for the prototype without CW multiplier\cite{10.1093/ptep/ptad146}.

    \subsection{Monitoring of the HV output voltage for the \SI{180}{\L} prototype detector}
\begin{figure}[tb]
        \centering
        \includegraphics[width=0.5\linewidth]{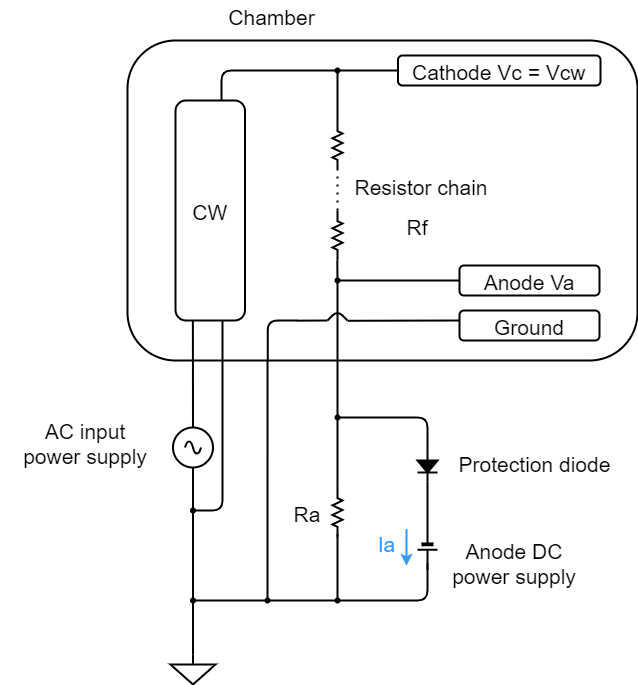}
        \caption{Schematic diagram of the HV supply to the \SI{180}{\L} prototype detector}
        \label{fig:cw_readout_HP180L}
    \end{figure}

    Figure \ref{fig:cw_readout_HP180L} shows the schematic diagram of the HV supplies to the \SI{180}{\L} prototype detector.
    When operating with the \SI{180}{\L} prototype detector, the CW output voltage is monitored with the voltage and current of the Anode DC power supply.
    A Matsusada HFR10-20N supplies the voltage across the ELCCs, which also is the TPC's anode voltage.
    This power supply has monitoring outputs for voltage and current and they are used to monitor the CW output voltage.
    
    The anode and cathode are at negative voltages, $-V_a$ and $-V_c$, respectively.
    The output voltage $V_{CW}$ of the CW multiplier is given by,
    \begin{equation}
      V_{CW} = V_c = \Bigl( 1 + \frac{R_f}{R_a} \Bigr) V_a - R_fI_a ,
      \label{eq:V_CW}
    \end{equation}
    Here, $R_a$ is the additional load resister (Fig.~\ref{fig:cw_readout_HP180L}) for the anode power supply, which is \SI{200}{\mega\ohm}, and $R_f$ is the resistance of the entire TPC resistance chain, which is \SI{8.1}{\giga\ohm}.
    The anode voltage $V_a$ and current $I_a$ are - as indicated above - taken from the monitoring outputs of the anode power supply.
    
\section{Operation as a HV supplier to the 180L detector}\label{sec:measurement}
\begin{figure}[tb]
    \begin{minipage}{0.5\linewidth}
        \centering
        \includegraphics[width=\linewidth]{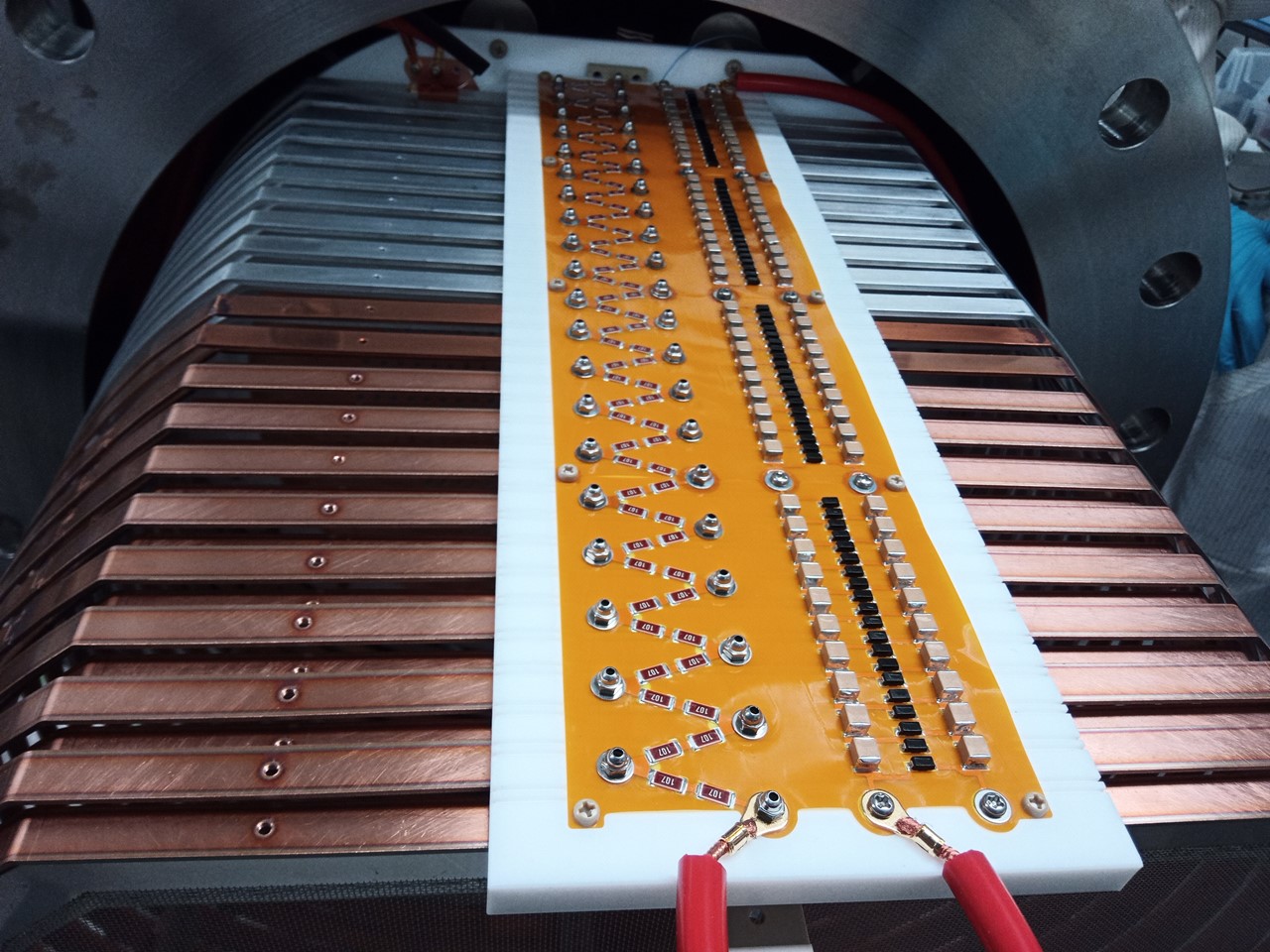}
    \end{minipage}
    \begin{minipage}{0.5\linewidth}
        \centering
        \includegraphics[width=\linewidth]{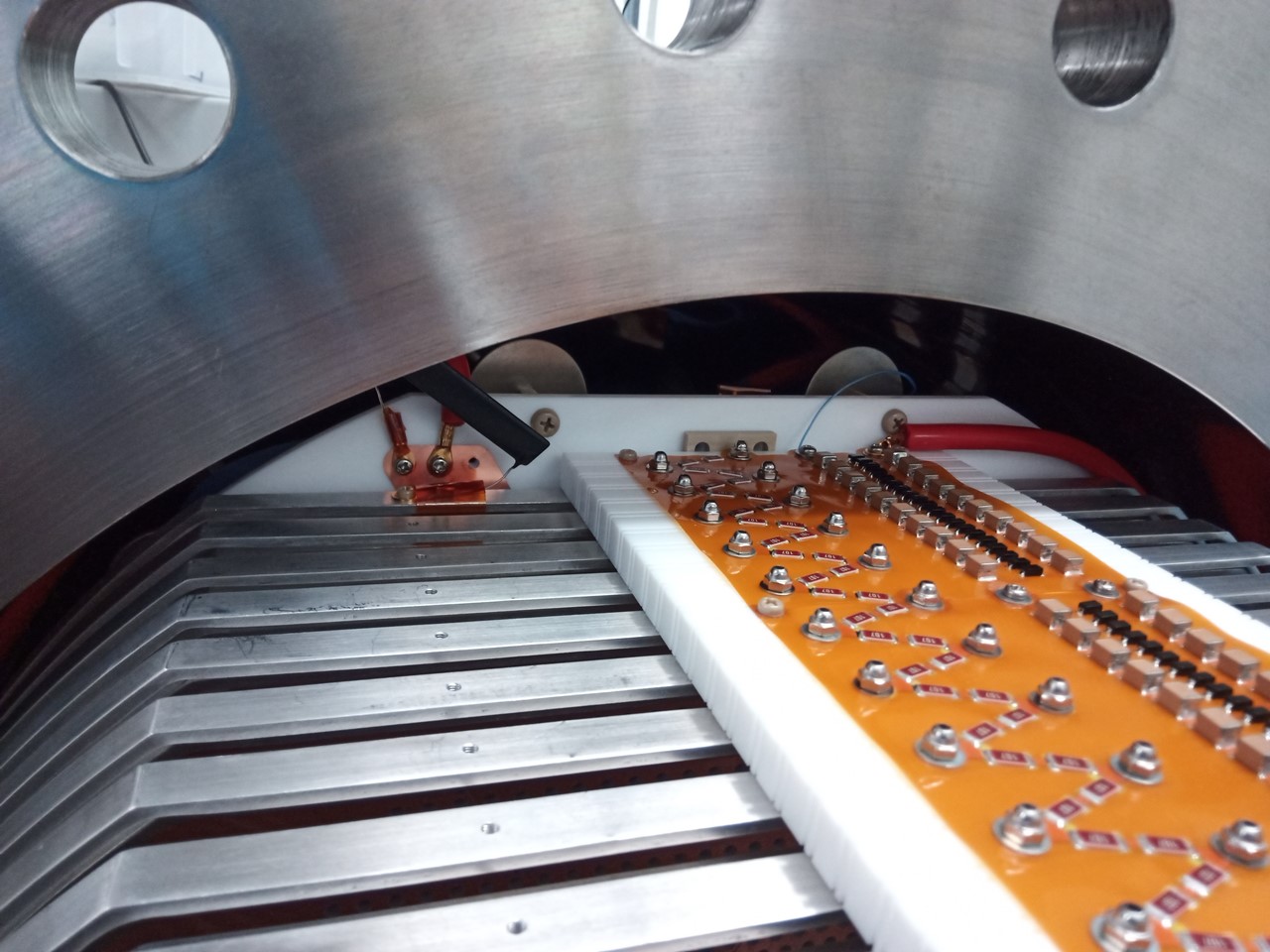}
    \end{minipage}          
    \caption{CW and resistor chain are mounted on a PTFE jig and installed on the field cage. The field cage and anode electrode are connected via a resistor FPD4D200MOHMF (black plate seen against the white PTFE at the end of the field cage).}
    \label{fig:cw_installed}
\end{figure}

Four CW multiplier and resistor chain boards were connected and fixed on a PTFE plate, which was installed on the flattened outside of the field cage as shown in Fig.~\ref{fig:cw_installed}.
To prevent electrical discharges, two grooves, oriented perpendicular to the drift direction, were machined at \SI{1}{\cm} intervals on the front, back, and side surfaces of the PTFE plate.
A stainless steel screw connects each electrode of the field cage to its corresponding stage in the resistor chain.
Although it is preferable to use coaxial cables to suppress AC pickup on the MPPC signal line, a silicon-rubber insulated single-wire cable was used for the input to the CW multiplier, because the use of coaxial cables causes oscillation of the AC amplifier due to the capacitive load of the cable.

The chamber was evacuated for about 5 days before introducing xenon gas.
The pressure reached \SI{4.5e-2}{\Pa} and the outgassing rate was \SI{1.5e-4}{\Pa.\m^3/\s}.
Then, the xenon gas was filled to about \SI{6.8}{\bar}.
Before applying HV gas purification was performed for 9 days.
The detector operating voltages were then adjusted while taking data on a trial basis.
After another 9 days to optimize detector operating conditions, data was taken from July 8, 2024 to September 13, 2024, with intervals in between.
The data used in the analysis were acquired for a total of 40 days.

As a discharge happened at around \SI{38}{\kV} during detector optimization, data-taking was conducted at the \SI{90}{\%} of the design value, \SI{34.3}{\kV}, which gives a \SI{90}{\V/\cm/\bar} drift electric field.
The applied voltage to the ELCC was also set at the \SI{90}{\%} of the design value, too.
The applied voltages were hence \SI{9.18}{\kV} for the anode and \SI{34.3}{\kV} for the cathode.
From a later inspection of the discharge tracks, it was inferred that the charge flowed from the cathode electrode and headed via the inner surface of the HDPE tube to the pressure vessel.

\begin{figure}[tb]
    \centering
    \includegraphics[width=0.6\linewidth]{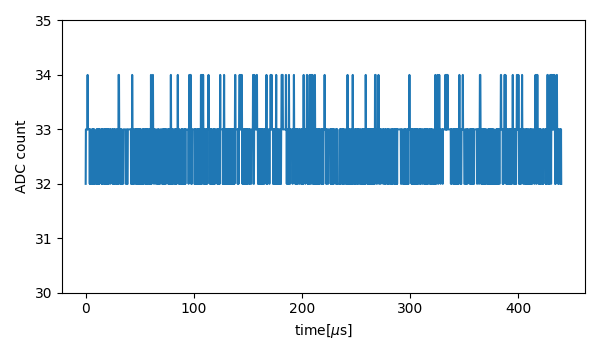}
    \caption{Example waveform of an ELCC channel without hits. Sampling rate is \mbox{\SI{5e6}{samples/s}} (\SI{0.2}{\us/sample}) . One ADC count corresponds to \SI{0.488}{\mV}. The input frequency of \SI{6}{\kHz} for the CW multiplier corresponds to \SI{167}{\us}.}
    \label{fig:sample_waveform}
\end{figure}

\begin{figure}[tb]
    \centering
    \includegraphics[width=0.5\linewidth]{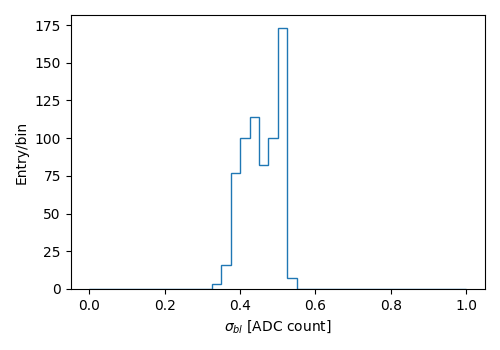}
    \caption{Distribution of the standard deviation of the baseline of the ELCC channels.}
    \label{fig:sigma_baseline}
\end{figure}

A sample waveform of an ELCC channel without hits during data acquisition is shown in Fig~\ref{fig:sample_waveform}.
The applied AC frequency of the CW multiplier is \SI{6}{\kHz} and the corresponding cycle is \SI{167}{\us}.
The baseline is stable within 1 to 2 ADC counts.
The distribution of the baseline standard deviations $\sigma_{bl}$ for all the ELCC channels is shown in Fig.~\ref{fig:sigma_baseline}.
The mean standard deviation is 0.46, while 0.45 when measured without high voltage applied.
The effect of baseline fluctuations within one ADC count on the energy resolution is small enough compared to other factors.
Therefore, the effect of AC pickup due to the CW multiplier is not a problem.

\begin{figure}[tb]
\begin{tabular}{cc}
    \begin{minipage}{0.5\linewidth}
    \centering
    \includegraphics[width=\linewidth]{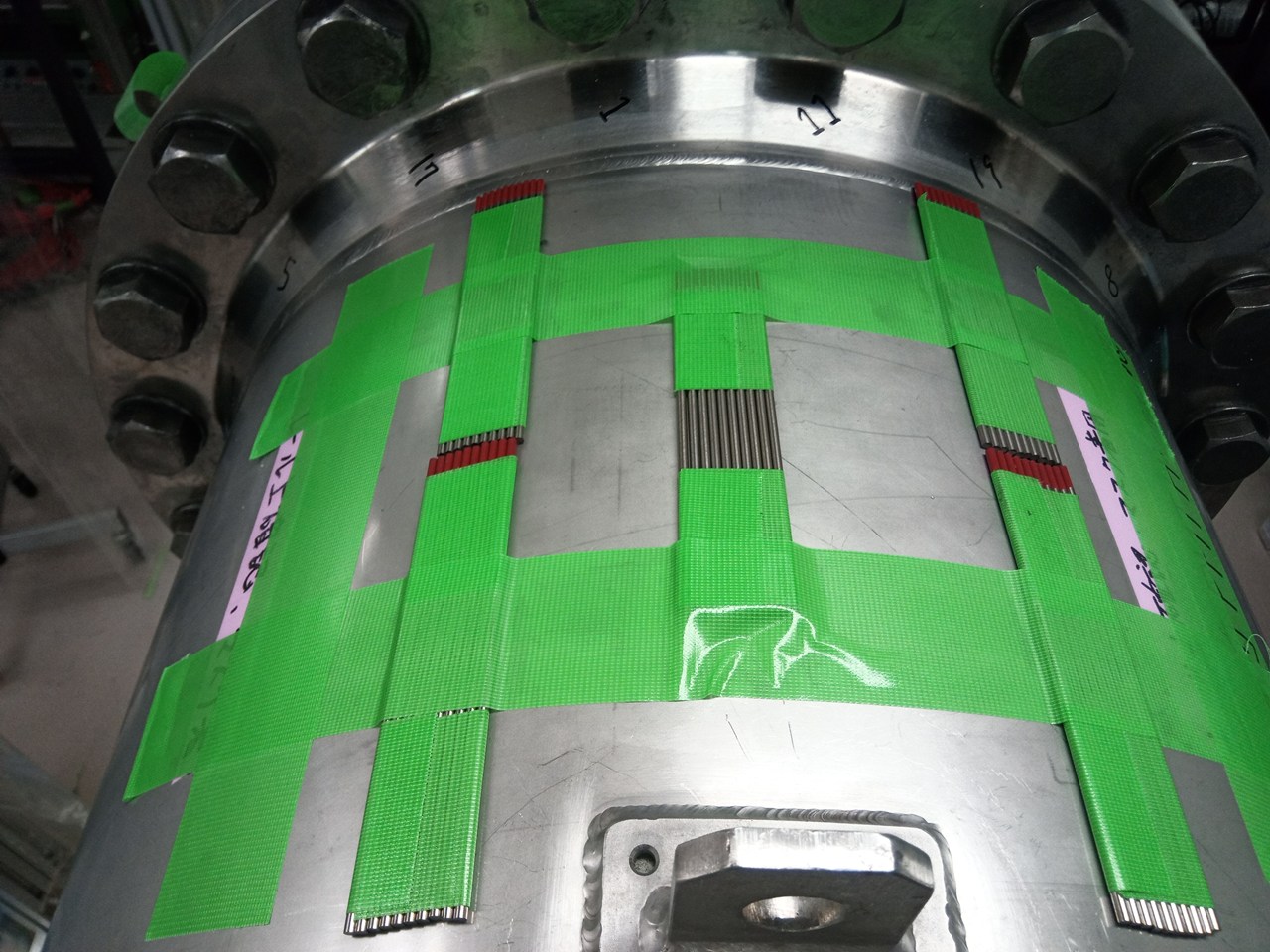}
    \end{minipage}
          
    \begin{minipage}{0.5\linewidth}
    \centering
    \includegraphics[width=\linewidth]{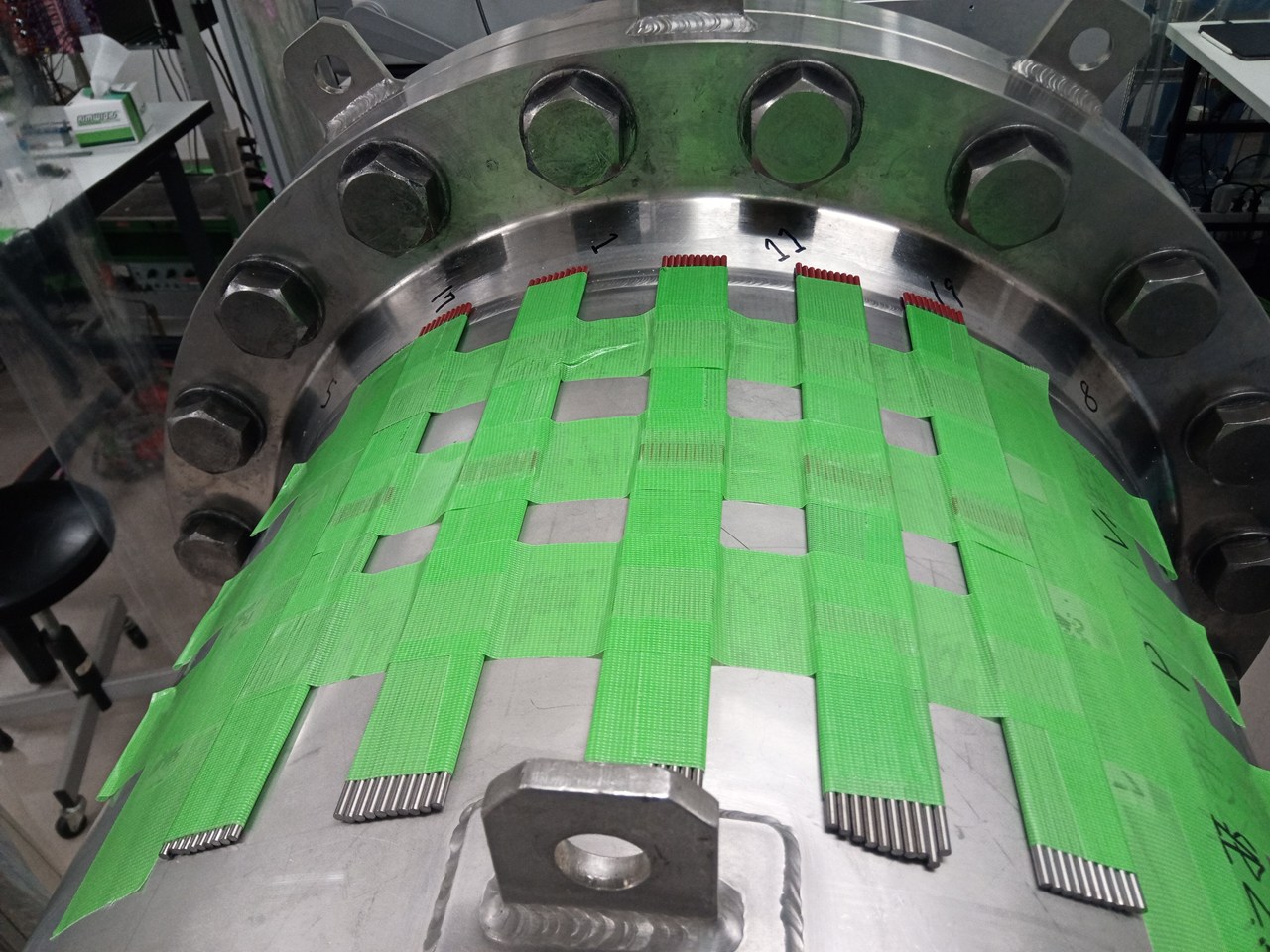}
    \end{minipage} \\
    
    \begin{minipage}{0.7\linewidth}
    \centering
    \includegraphics[width=\linewidth]{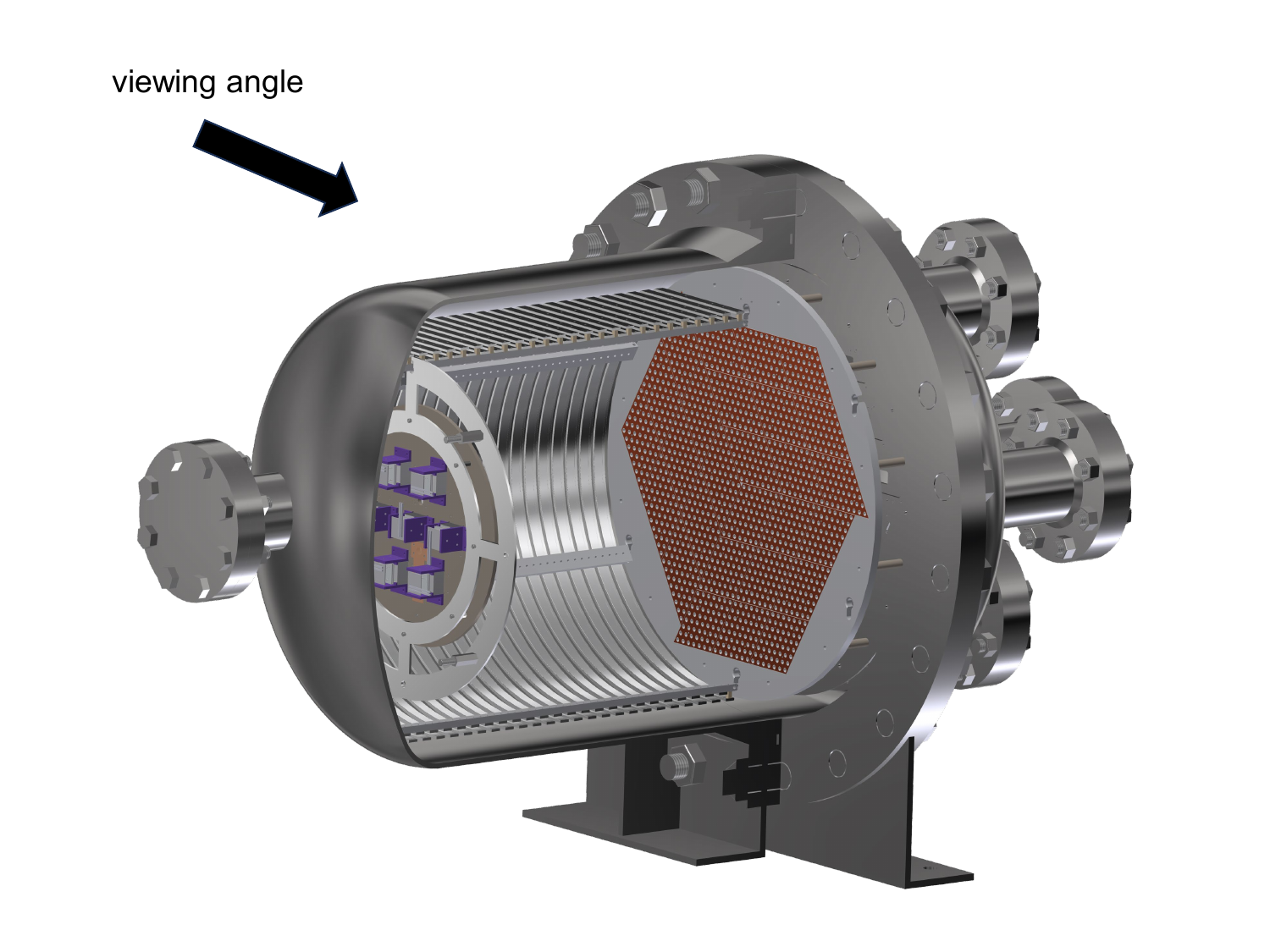}
    \end{minipage}
\end{tabular}
\caption{Installation of \SI{1}{\kg} (left) and \SI{2}{\kg} (right) of thorium-doped tungsten rods on the chamber. Ten rods (\SI{200}{\g}) are bound together with curing tape and attached to the upper side of the circumference of the cylindrical part of the pressure vessel. The sectional view of the \SI{180}{\L} prototype detector on the bottom figure shows the viewing angle.}
\label{fig:source_distribution}
\end{figure}

We used thorium-doped tungsten rods as a gamma-ray source.
This is a commercial product for welding and contains \SI{2}{\%} thorium and hence contains $^{208}\mathrm{Tl}$ in the thorium series.
A $^{208}\mathrm{Tl}$ nucleus emits a gamma ray of \SI{2615}{\keV}, which is close to the energy of $\beta$-rays from the $^{136}\mathrm{Xe}~0\nu\beta\beta$, \SI{2458}{\keV}.
The weight of the thorium-doped tungsten rods used was \SI{1}{\kg} at the beginning and later doubled.
The intensity is \SI{80}{kBq} and \SI{160}{kBq} respectively.
The rods are attached to the upper side of the circumference of the cylindrical part of the pressure vessel (Fig.~\ref{fig:source_distribution}).

\begin{figure}[tb]
    \centering
    \includegraphics[width=0.6\linewidth]{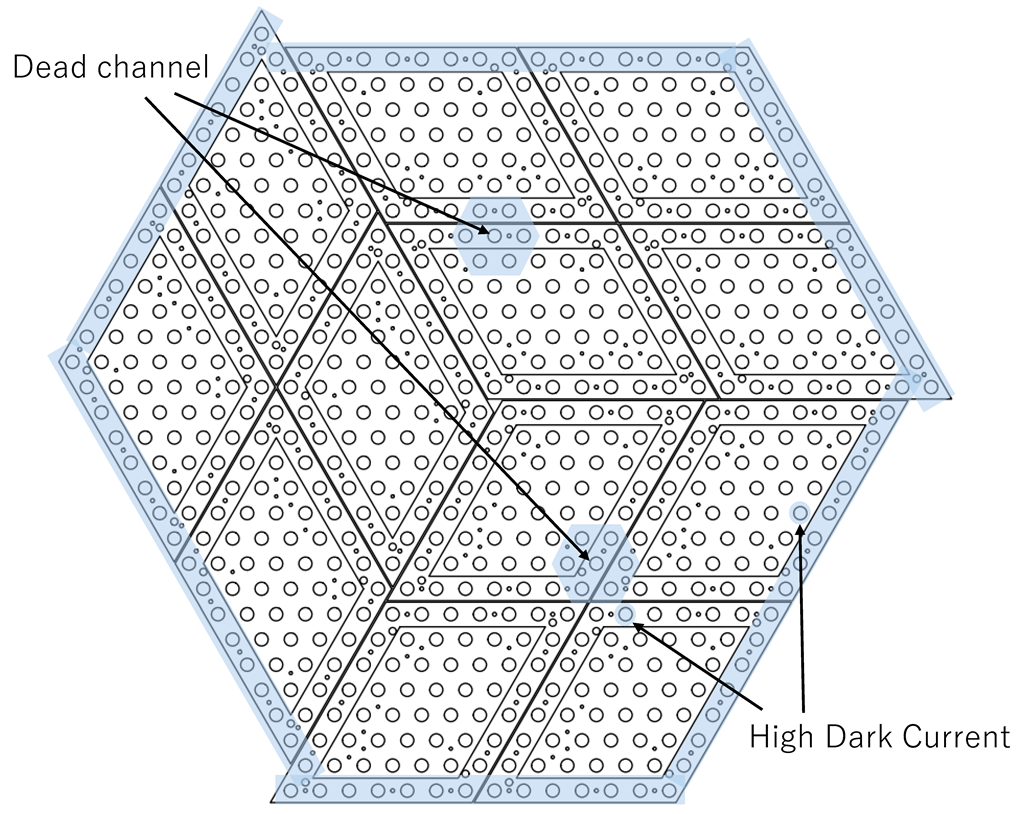}
    \caption{Configuration of veto channels. The blue-shaded channels were assigned to veto.}
    \label{fig:Veto_channel}
\end{figure}

The outer cells of the ELCC plane were set as veto to select fully contained events.
There were two dead channels and two high dark current channels (see Fig.~\ref{fig:Veto_channel}).
The high dark current channels were also added to the veto, as were the channels surrounding dead channels.

\begin{figure}[tb]
    \centering
    \includegraphics[width=0.95\linewidth]{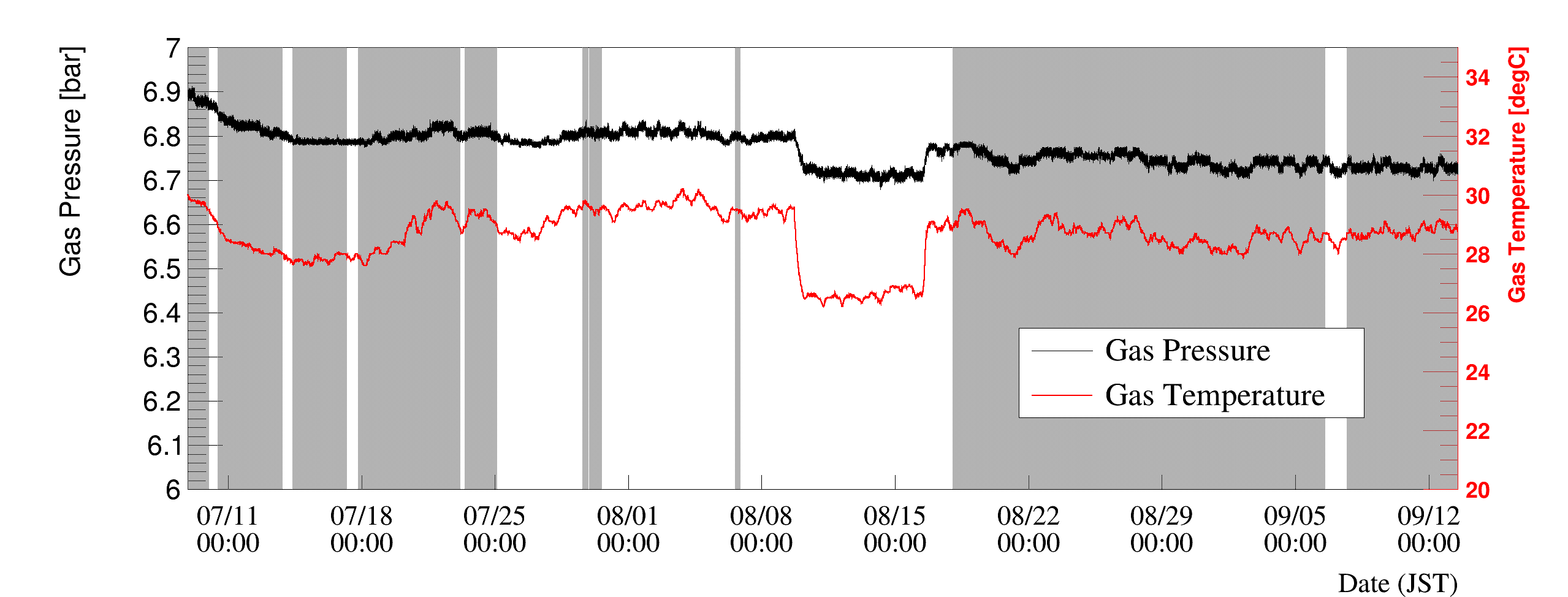}
    \includegraphics[width=0.95\linewidth]{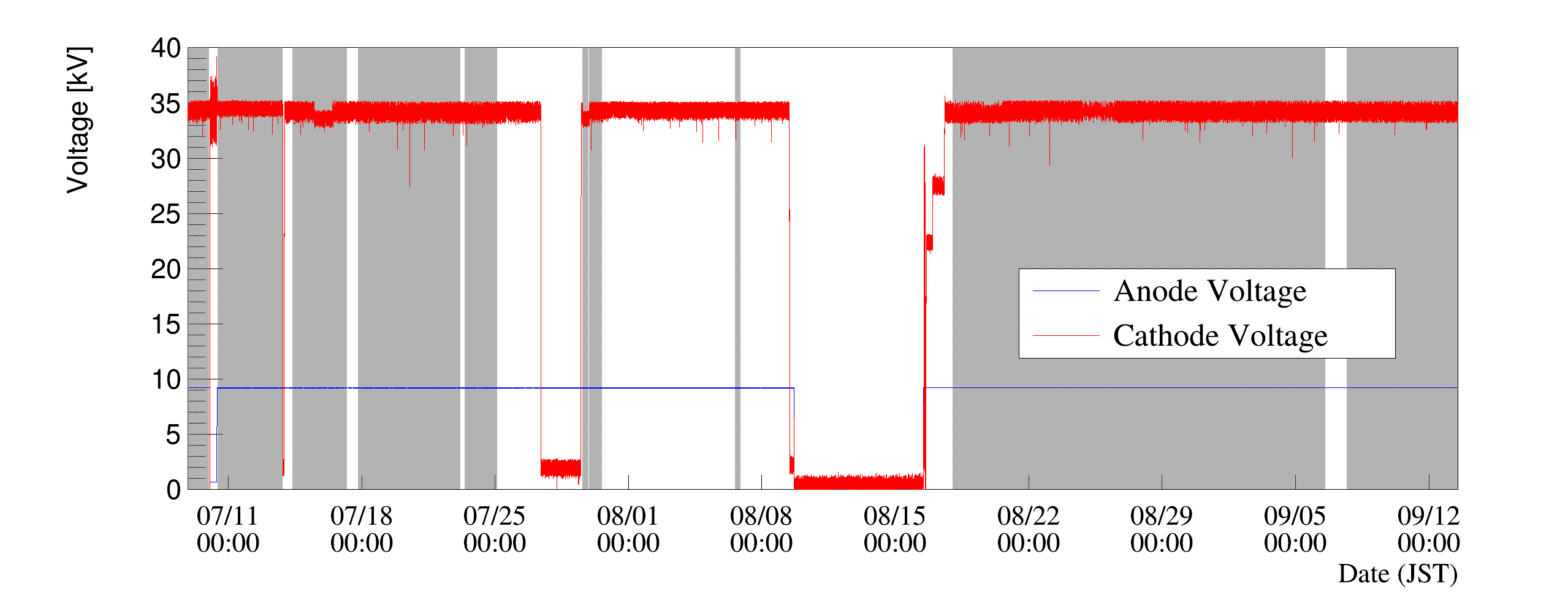}
    \caption{Trends of the gas conditions (upper), and high voltages (lower). The gray-shaded areas are data-taking periods. The drops in the anode and cathode voltage from August 9 to 16 correspond to manual shutdowns, while other voltage drops (including drops on July 10) correspond to discharges.}
    \label{fig:monitor}
\end{figure}

\begin{figure}[tb]
    \centering
    \includegraphics[width=0.9\linewidth]{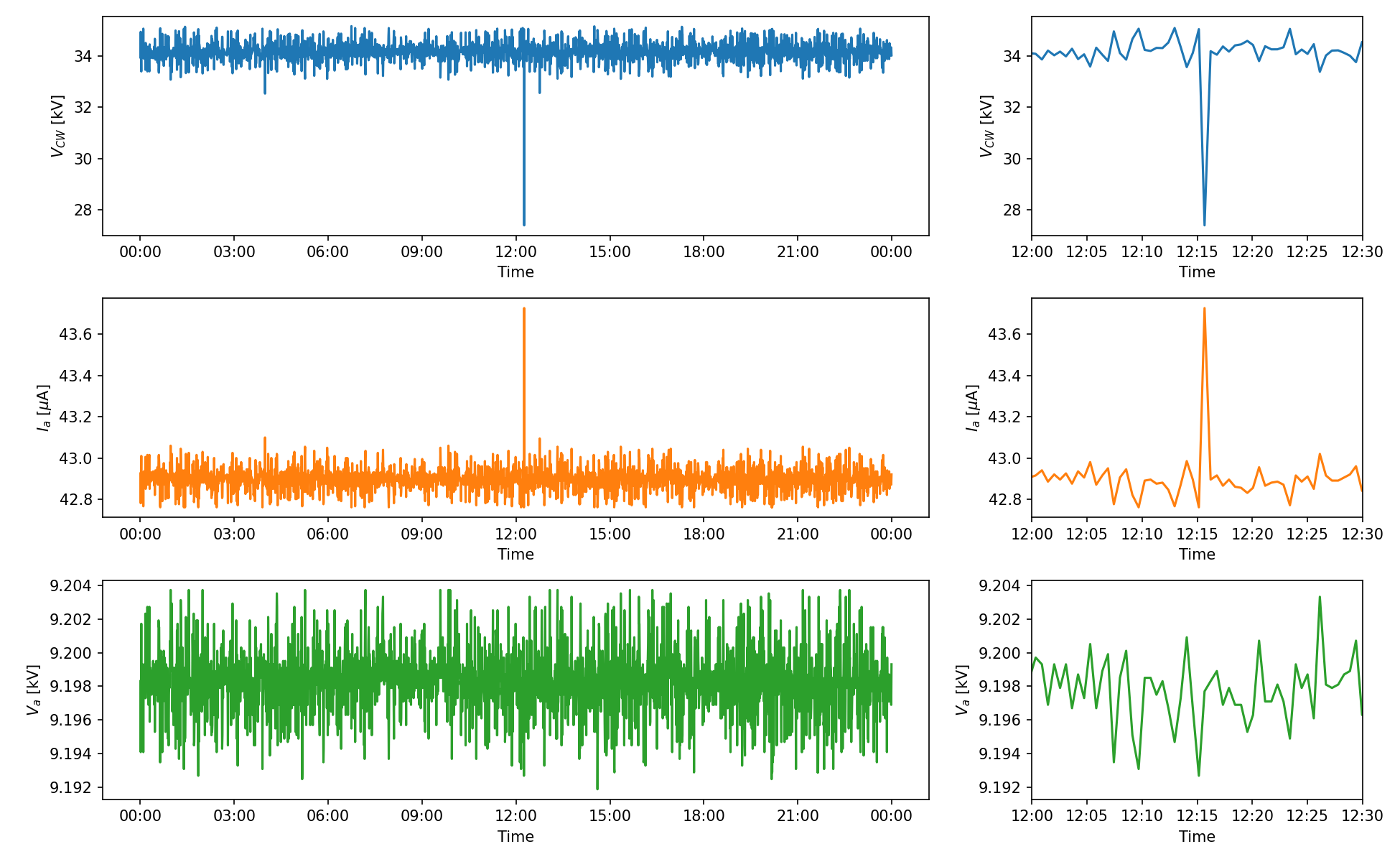}
    \caption{Trend of the CW output voltage (top), anode current (middle), and anode voltage (bottom) on July 20. The figures on the right are enlargements of the range from 12:00 to 12:30. One bin is equivalent to a sampling interval of 33 seconds.}
    \label{fig:monitor_20240720}
\end{figure}

Various monitored quantities during the data taking period are shown in Fig.~\ref{fig:monitor}.
The sampling interval for these monitored quantities is 33 seconds.
The CW output voltage has a temporal variation of about \SI{1}{\kV} and can drop several kV momentarily.
As an example, the CW output voltage, anode current, and anode voltage on July 20 are shown in Fig.~\ref{fig:monitor_20240720}.
The CW output voltage fluctuates by \SI{0.4}{\kV} around \SI{34.2}{\kV} during this period.
As the cathode voltage is calculated from the monitored anode current and voltage, the sharp CW voltage drop of \SI{7}{\kV} observed in one bin near noon in Fig.~\ref{fig:monitor_20240720} likely is an artifact from a discharge in the CW or ELCC.
The frequency of such a momentary voltage drop was less than a few times per day and did not cause DAQ outages.
In another incident on July 27, the CW output voltage dropped to about \SI{2}{\kV}.
The anode voltage of \SI{9.18}{\kV} remained applied, and the voltage drop in the resistor chain was probably caused by reverse current flow through the diodes in the CW multiplier.
This voltage drop was because the voltage input to the CW multiplier was stopped by an interlock that was triggered by such a discharge.
Moisture content was also monitored using a dew point meter and was below the lower limit of the meter's range for most of the measurement period.
This means that the moisture content was less than 0.05 ppm.

There are two kinds of triggers, the fiducial trigger and the whole trigger, for the data acquisition of the ELCC signal\cite{10.1093/ptep/ptad146}.
The whole trigger is issued when the summed signal height of all channels exceeds a threshold.
The fiducial trigger is issued similarly except that the veto channels must have no hits.
The whole trigger is to collect the data of xenon $\mathrm{K}_\alpha$ characteristic X-ray (\SI{29.68}{\keV}) for calibration, while the fiducial trigger is to accumulate higher energy events.
For the measurements reported below, the threshold of the fiducial trigger was set to about \SI{500}{\keV} and the whole trigger to slightly above the baseline.
The fiducial trigger rate for this threshold was \SI{10.5}{\Hz}.
To reduce data volume, the whole trigger was set to be issued only once every \SI{50000}{} times. 

These triggers and veto signals are issued with a trigger board, Hadron Universal Logic module\cite{R.Honda2016}.
According to these triggers, AxFEB acquires ELCC waveforms with a time width of \SI{440}{\us}, including \SI{40}{\us} before the trigger, and for PMT signals, the waveform digitizer stores PMT waveforms with a window of \SI{900}{\us} with pre-trigger region of \SI{95}{\%} of the recording window.
The ELCC waveforms and the PMT waveforms obtained from a single trigger constitute data for one event.

\section{Analysis}\label{sec:analysis}
The fiducial trigger conditions were finalized on July 8, 2024, and data was taken with a stable detector until September 13, 2024.
We used the data from this period for analysis.
Here, we briefly describe the method for estimating the energy of each event from the obtained data.
For a detailed description of the analysis method, see \cite{10.1093/ptep/ptad146}.

  \subsection{ELCC waveform analysis}
  From the waveform of each ELCC channel, hits were identified with a certain threshold from the baseline.
  Photon counts of each hit were calculated using the MPPC gain.
  Since MPPC pixels need a recovery time to restore the bias voltage after the charge is released by photon detection, MPPCs have a non-linear output for high incident light intensity.
  This MPPC non-linearity was corrected using the recovery times measured in advance\cite{11.1093/ptep/ptaa030}.
  From the hits clusters were formed by grouping those that were spatially and temporally adjacent for each event.
 
  Photon counts of each cluster or each event were obtained by summing up those of hits in the cluster after the gain of the EL process in each ELCC cell (hereafter, EL gain) was corrected for.
  The EL gain was defined as the average detected photon count for one ionization electron.
  The EL gains were different cell-by-cell due to machining inaccuracy and photon detection efficiency differences of MPPCs.
  To determine the correction factor, peaks of xenon $\mathrm{K}_{\alpha}$ clusters detected throughout the measurement were used.

  The average EL gain in this measurement was found to be 11.5, which is smaller than the 12.5 in the previous study\cite{10.1093/ptep/ptad146}.
  This presumably is because the number of photons reaching the MPPC decreased due to the change of the geometry of ELCC cell; \SI{4.5}{\mm} open hole to \SI{4.1}{\mm} tapped hole.
    
  \subsection{PMT waveform analysis}
\begin{figure}[tb]
    \centering
    \includegraphics[width=0.6\linewidth]{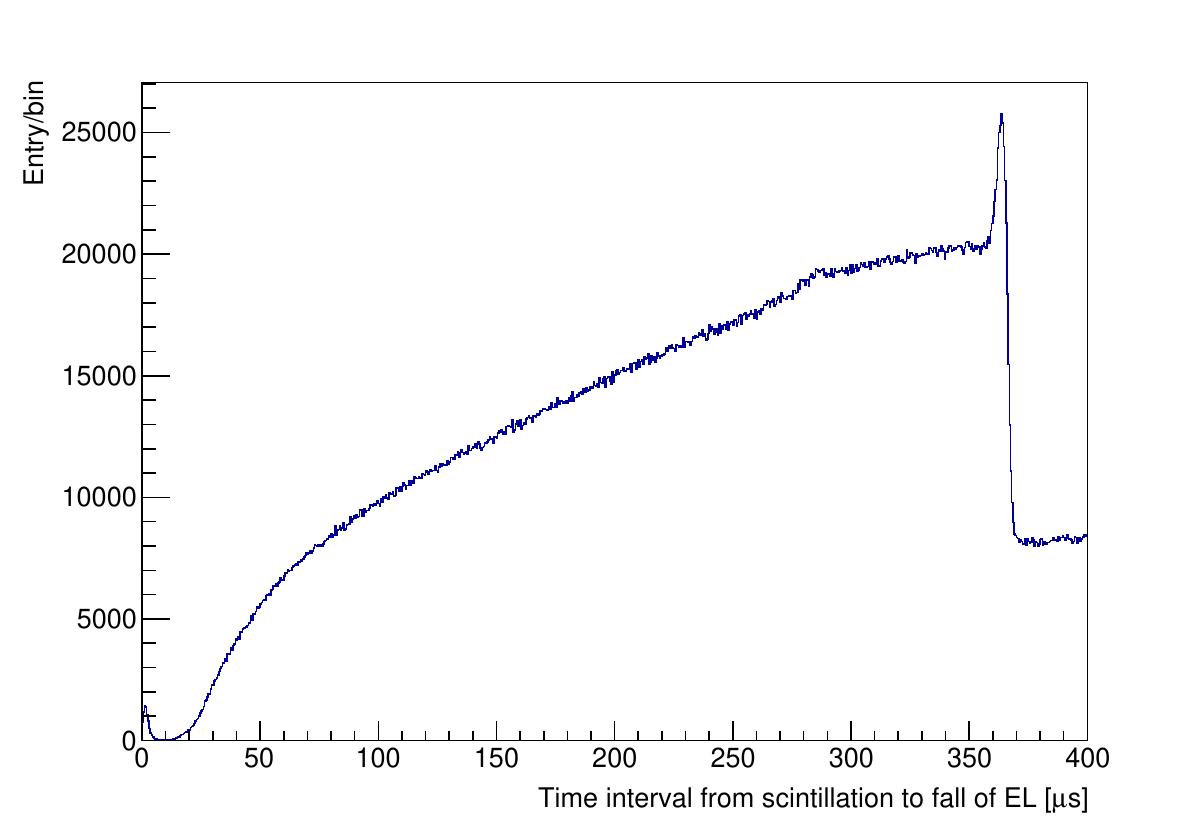}
    \caption{Distribution of the time intervals between scintillation and the end of the ELCC signal. The peak at \SI{363}{\us} is formed by the events across the cathode plane.}
    \label{fig:drift_time}
  \end{figure}

  To select primary scintillation photon candidates, hits were detected from the PMT waveform with a certain threshold from baseline.
  Because hits included EL light from ELCC cells as well as primary scintillation light, hits with a width of less than \SI{400}{\ns} and at least \SI{1}{\us} away from other hits were selected as primary scintillation light candidates.
  Of these hits, when two or more PMTs had hits within \SI{100}{\ns}, they were considered as primary scintillation light hit clusters.
  Events with multiple clusters were rejected.
  The distribution of the time intervals between the primary scintillation and the end of the ELCC signal is shown in Fig.~\ref{fig:drift_time}.
  The peak at \SI{363}{\us} is formed by the events crossing the cathode of the field cage, $z=40~\mathrm{cm}$.
  From this, the drift velocity of ionization electrons was derived to be \SI{1.10}{\mm/\us}.
  The z-positions of ELCC hits were reconstructed using this drift velocity.

  There are additional selections to avoid timing mismatches, see the previous study\cite{10.1093/ptep/ptad146} for details. 

  \subsection{Fiducial volume cut and overall corrections}
  Using the information from ELCC and PMT, an additional fiducial volume cut was applied.
  This fiducial volume cut was applied to reject events with hits in the veto channel and restrict the z-position to the $\SI{8}{cm}<z<\SI{36}{cm}$ region.
  But there were small photon count clusters due to a few electrons that disturbed the fiducial volume cut.
  These electrons were considered to be produced by the photoelectric effect caused by EL light or by the release of electrons attached to impurities, and eliminated from events before the fiducial volume cut.
  Then, following overall corrections were applied.

    \subsubsection{Correction for temporal variations of photon counts}
\begin{figure}[tb]
        \begin{minipage}{0.5\linewidth}
          \centering
          \includegraphics[width=\linewidth]{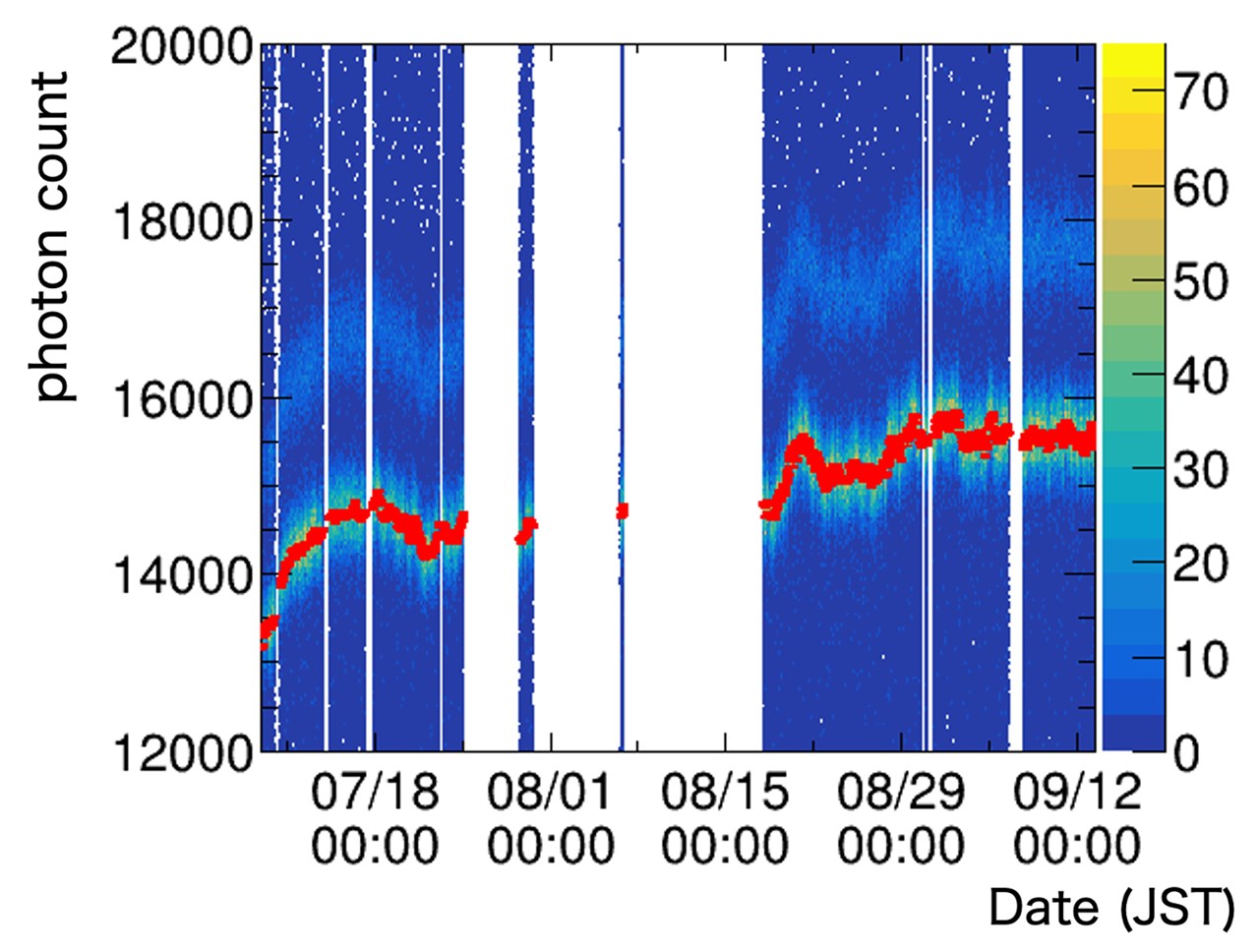}
        \end{minipage}
        \begin{minipage}{0.5\linewidth}
          \centering
          \includegraphics[width=\linewidth]{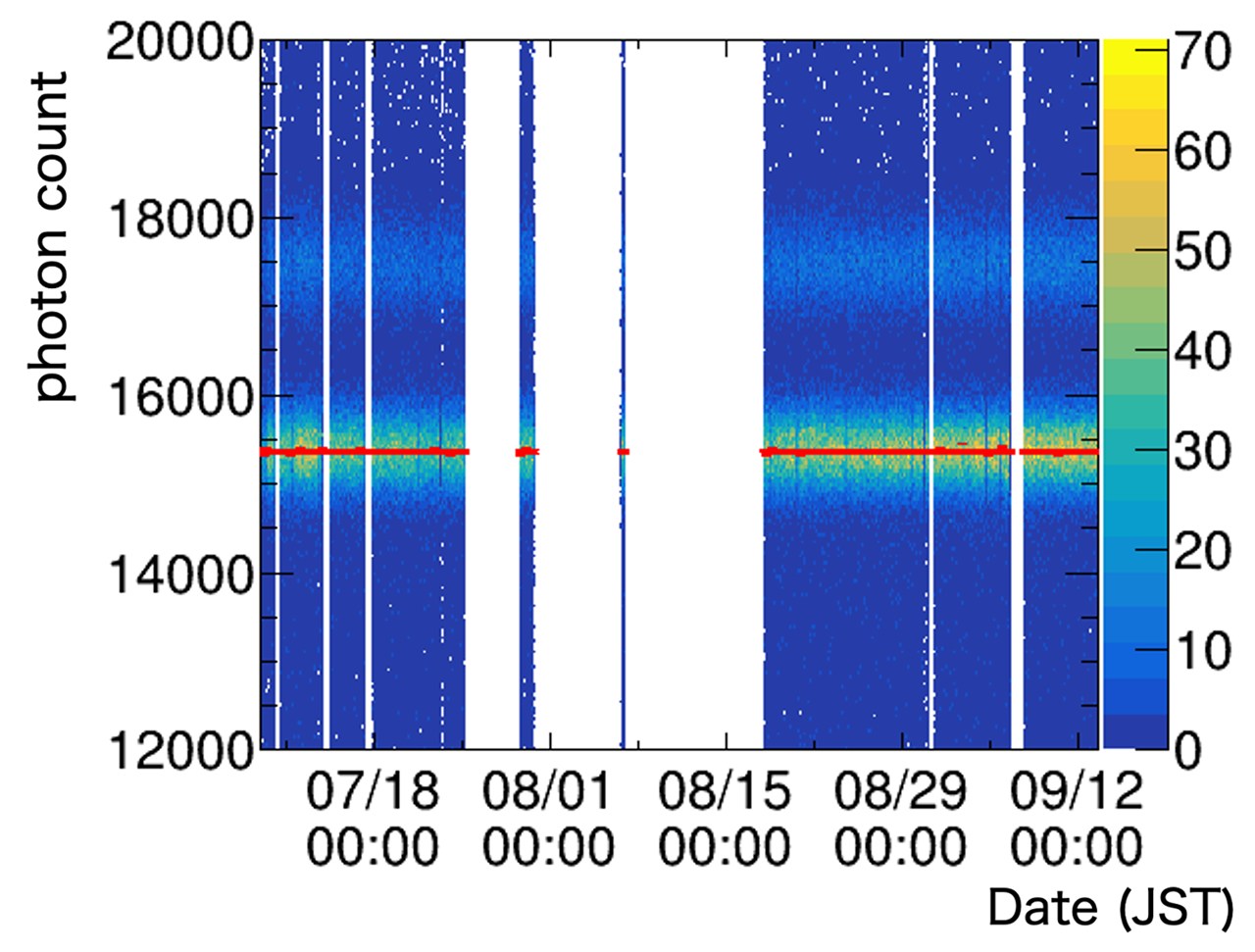}
        \end{minipage}          
        \caption{Temporal variation of the photon count of xenon $\mathrm{K}_{\alpha}$ characteristic X-rays before (left) and after (right) correction. Red lines represent the xenon $\mathrm{K}_{\alpha}$ peak position and its fitting error.}
        \label{fig:tcor}
      \end{figure}

      EL light intensity varied over time.
      Pressure variation with gas temperature changes the reduced electric field strength, which in turn changes the EL yield.
      Also, because impurities absorb EL light, EL light intensity varies with gas purity.
      Therefore, corrections were applied for 30 minute intervals, using the xenon  $\mathrm{K}_{\alpha}$ peaks.
      Figure \ref{fig:tcor} shows the time evolution of the photon counts from xenon $\mathrm{K}_{\alpha}$ peaks before and after correction.
      The photon counts corresponding to the xenon $\mathrm{K}_{\alpha}$ peaks were corrected to be uniform in time.

    \subsubsection{Correction of the $z$-dependence of the photon count}
\begin{figure}[tb]
        \centering
        \includegraphics[width=0.6\linewidth]{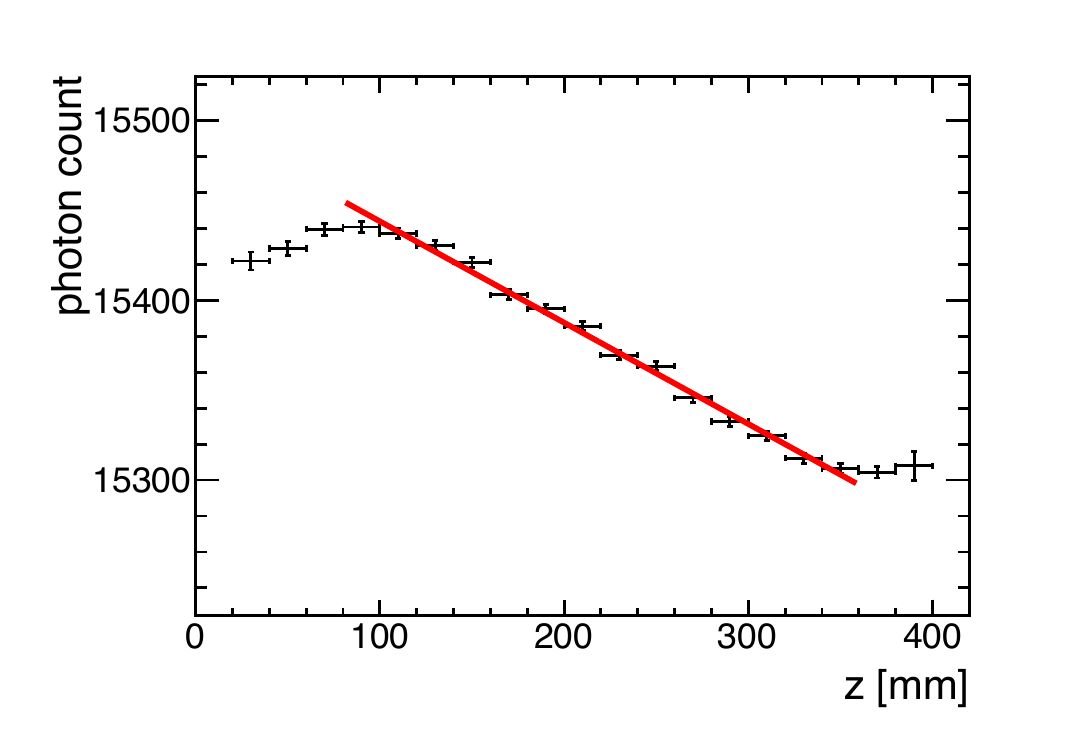}
        \caption{Dependence of the photon counts of xenon $\mathrm{K_\alpha}$ clusters on the $z$-position. The red line shows a linear fit.}
        \label{fig:z_dependence}
      \end{figure}

      There was an attenuation of ionization electrons during drift due to attachment to impurities.
      The photon count of xenon $\mathrm{K_\alpha}$ clusters is shown in Fig.~\ref{fig:z_dependence} as a function of the z position, from which the attenuation length was determined to be \SI{27500+-1020}{\mm}.
      This corresponds to an electron lifetime of \SI{25.0+-0.9}{\ms}.
      The photon counts for every waveform sample were corrected using this attenuation length.
      Deviations were observed for the smallest and the largest drifts, but the cause is not known.
      A possible reason is the mis-reconstruction of the z-position.
      Regarding the deviation on the smallest drifts, another possible reason is the insufficient MPPC non-linearity correction.
      On the smallest drifts, diffusion of ionization electrons during drift is suppressed, resulting in a relatively higher photon counts per unit time and very large MPPC non-linearity.
      The MPPC signal is shaped by analog filters and digitized at a finite sampling rate\cite{9072179}, and the correction may be inaccurate for such a case.

    \subsubsection{Overall fine-tuning for the non-linearity of MPPCs}\label{sec:fine-tuning}
\begin{figure}[tb]
        \begin{tabular}{cc}
          \begin{minipage}{0.5\linewidth}
            \centering
            \includegraphics[width=\linewidth]{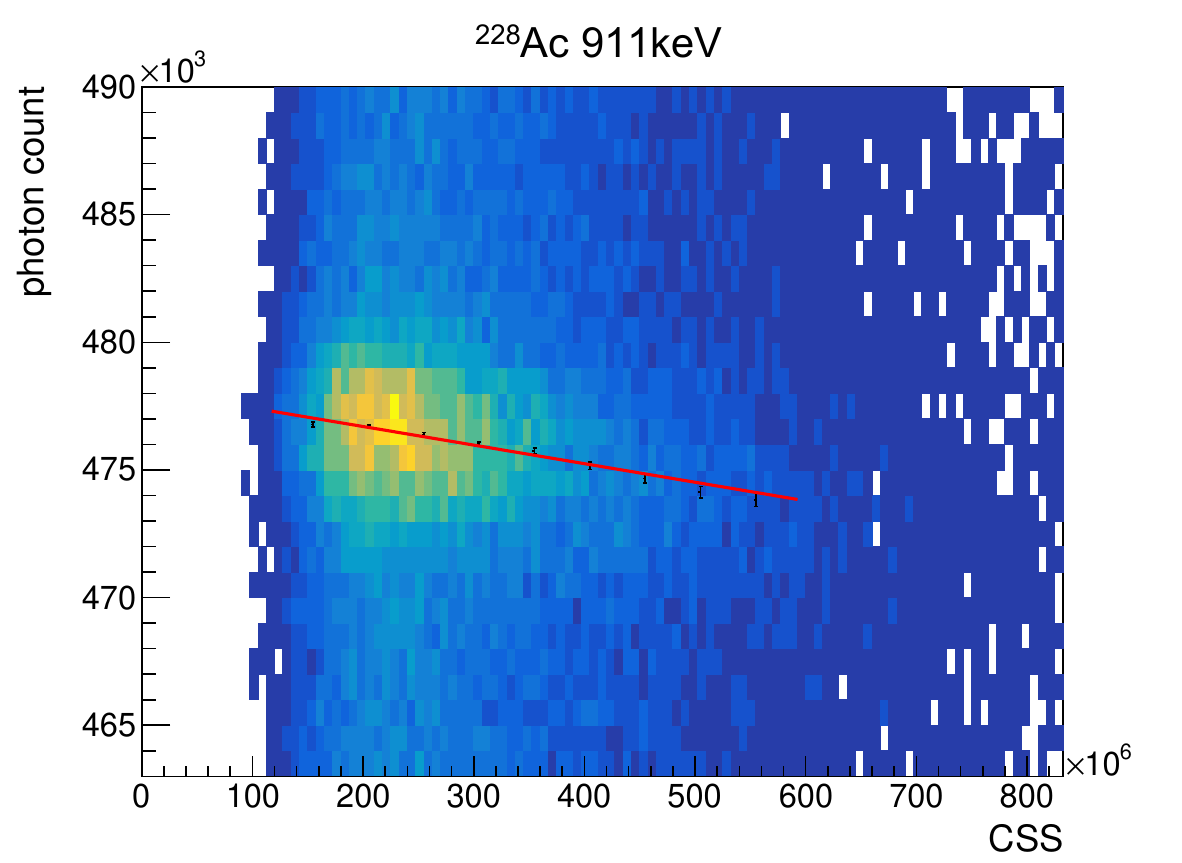}
          \end{minipage}
          
          \begin{minipage}{0.5\linewidth}
            \centering
            \includegraphics[width=\linewidth]{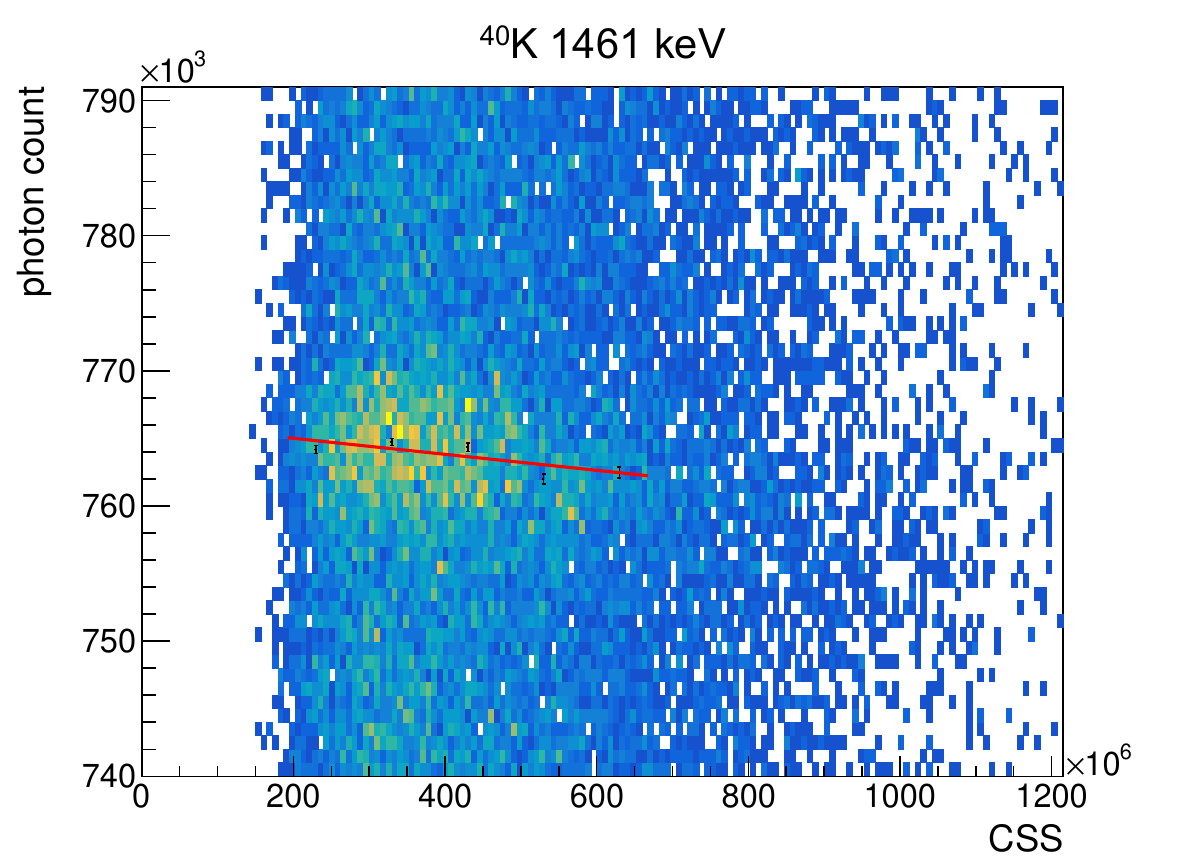}
          \end{minipage} \\
          
          \begin{minipage}{0.5\linewidth}
            \centering
            \includegraphics[width=\linewidth]{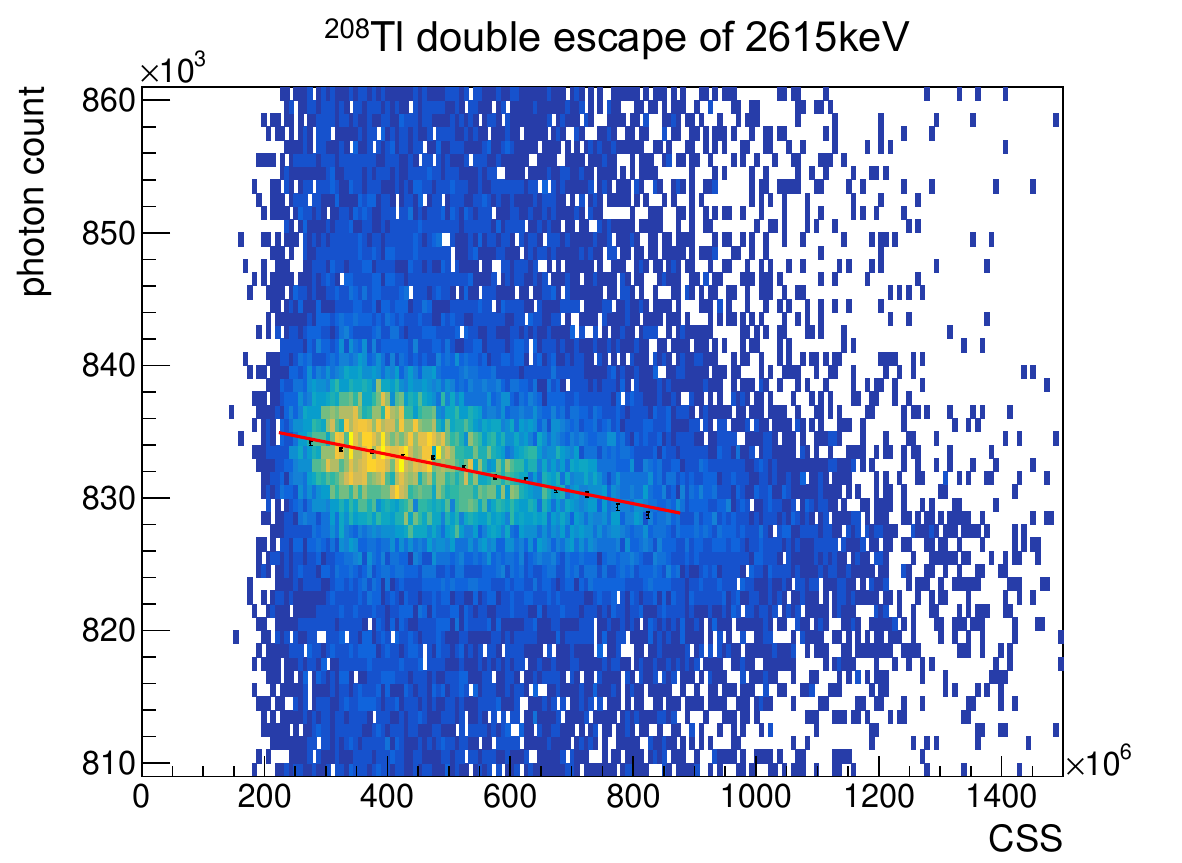}
          \end{minipage}        

        \end{tabular}
        \caption{Photon counts vs. CSS of the photopeak of \SI{911}{\keV} gamma rays from $^{228}$Ac (top left), \SI{1461}{\keV} gamma rays from $^{40}$K (top right) and double escape peak of \SI{2615}{\keV} gamma rays from $^{208}$Tl (bottom).}
        \label{fig:css}
      \end{figure}

      The effective recovery time of an MPPC can vary from the pre-measured one due to temperature changes, etc.
      If there remains an overall difference between the effective and the independently calibrated MPPC non-linearity, the effect appears as a linear relation between the photon counts and corrected squared sum, CSS: $\sum_ir^i\left(N_\mathrm{rec}^i\right)^2$ of events as follows\cite{10.1093/ptep/ptad146}.
      \begin{equation}
        \sum_i r^iN^i_{\mathrm{rec}} - N_{\mathrm{true}} \sim \Delta k \sum_ir^i(N^i_{\mathrm{rec}})^2
        \label{eq:photon_css}
      \end{equation}
      Here, i runs over every sampling of the waveforms of every hit channel, $r^i$ is the correction factor other than the MPPC non-linearity, $N_\mathrm{rec}^i$ are the photon counts after the MPPC non-linearity correction, and $N_{\mathrm{true}}$ is the true total photon count of the event.
      $\Delta k = k - k'$ is the difference between the true and calibrated MPPC non-linearity.
      The true (calibrated) non-linearity parameter $k^{(')}$ is expressed as $k = \tau^{(')} / (\Delta t \cdot N_{\mathrm{pixel}})$ using the true (calibrated) MPPC recovery time $\tau$, sampling interval $\Delta t$, and the number of MPPC pixels $N_{\mathrm{pixel}}$\cite{10.1093/ptep/ptad146}.
      We used the photopeaks of the \SI{911}{\keV} gamma rays from $^{228}\mathrm{Ac}$, \SI{1461}{\keV} gamma rays from $^{40}\mathrm{K}$, and the double escape peak of \SI{2615}{\keV} gamma rays from $^{208}\mathrm{Tl}$ to evaluate this effect.
      Distributions of photon counts vs. CSS for these peaks are shown in Fig.~\ref{fig:css}.
      As described in \cite{10.1093/ptep/ptad146}, the observed slopes indicate the overall bias of the MPPC non-linearity correction through the recovery time of MPPCs.
      The MPPC non-linearity correction, EL gain correction, correction for the temporal variation of the photon counts and correction of the $z$-dependence of the photon counts were then repeated with the recovery times shifted by this overall bias.

\section{Detector performance}
  
  \subsection{Energy resolution}
\begin{figure}[tb]
      \centering
      \includegraphics[width=0.8\linewidth]{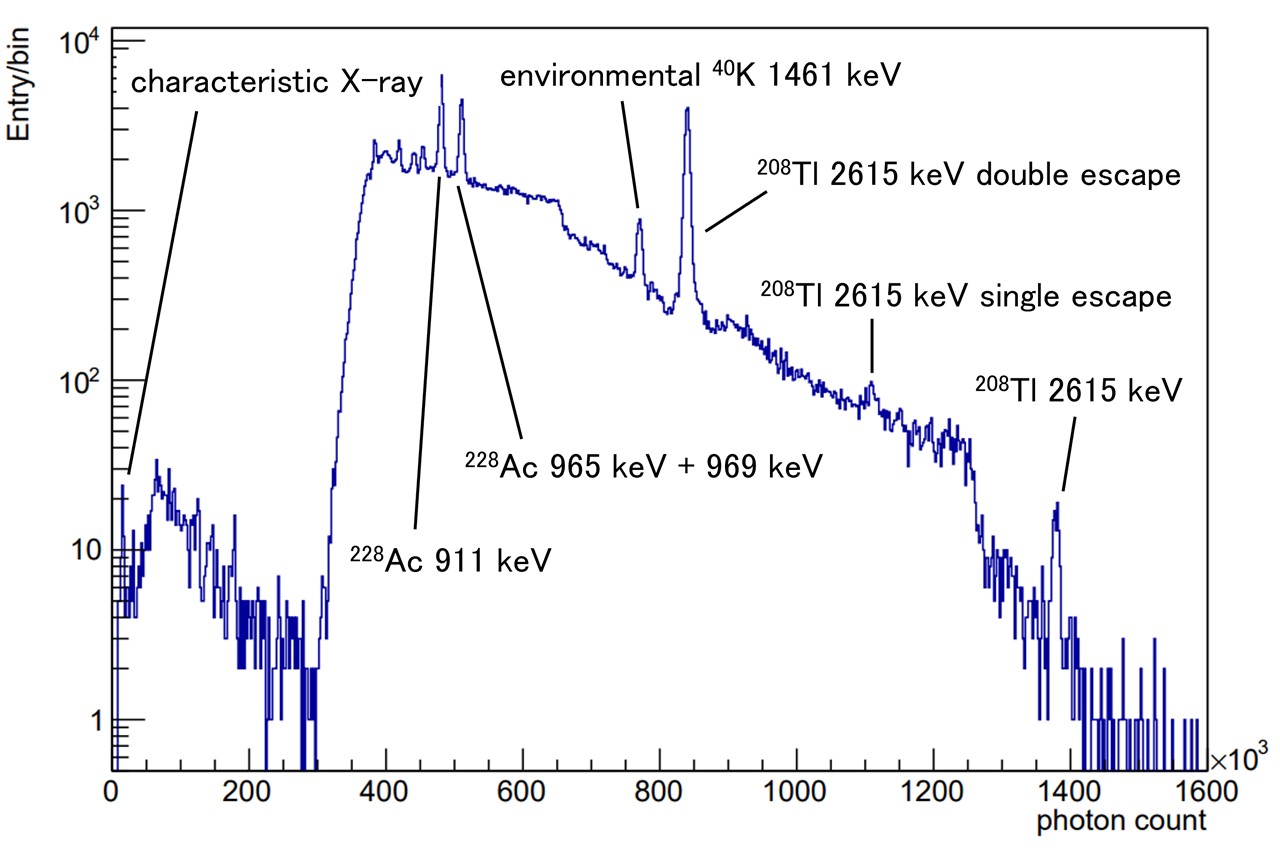}
      \caption{Photon count spectrum after applying all corrections and cuts. The drop in the spectrum below \SI{3e5}{} photons is due to the downsampling that suppresses the low energy data rate.}
      \label{fig:photon_spectrum}
    \end{figure}

\begin{figure}[tb]
      \centering
      \includegraphics[width=0.8\linewidth]{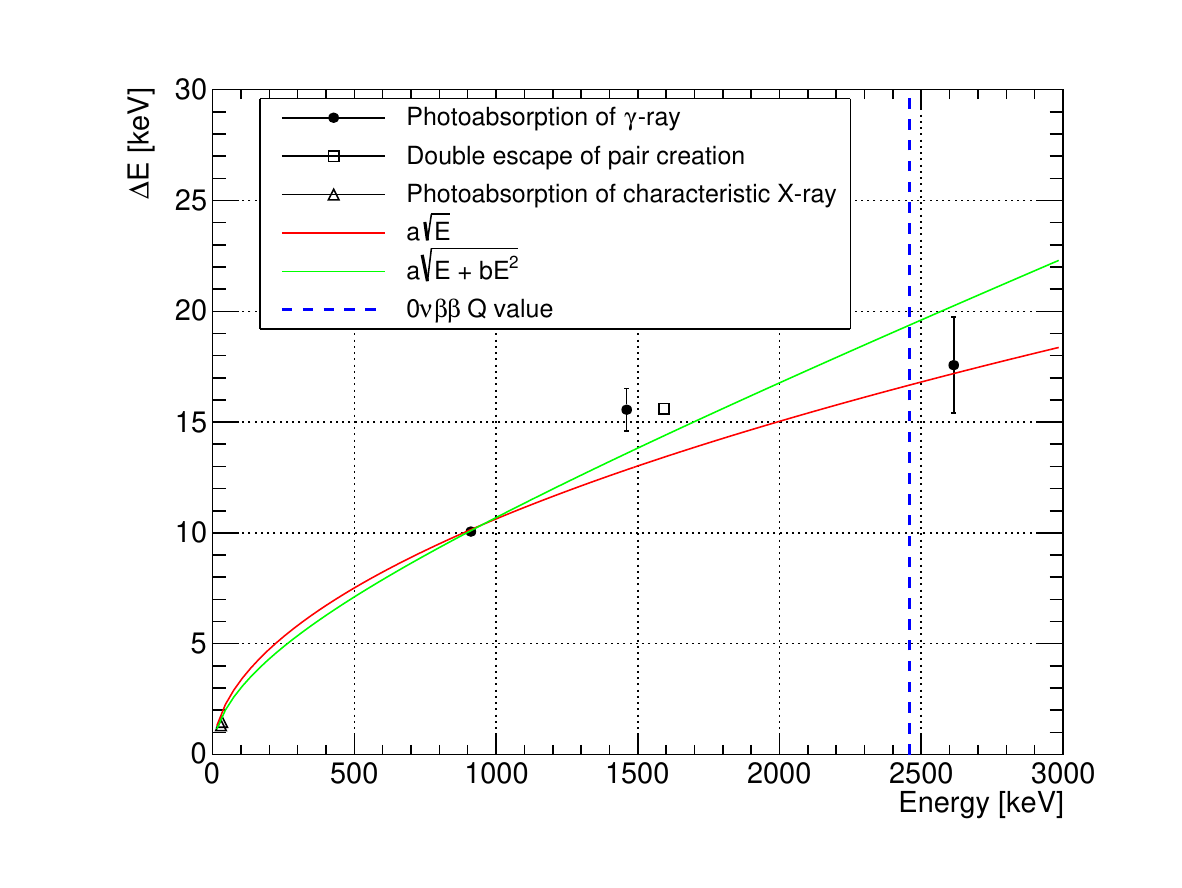}
      \caption{Interpolation of the energy resolution to the Q value.}
      \label{fig:resolution_fit}
    \end{figure}

    The EL photon count spectrum obtained during the data acquisition period described in Sec.~\ref{sec:measurement} is presented in Fig.~\ref{fig:photon_spectrum}.
    The $^{208}\mathrm{Tl}$ \SI{2615}{\keV} gamma ray peak and the single escape peak are clearly seen.
    Each peak of the spectrum was fitted with a combination of a Gaussian and a linear function, and the results are summarized in Table~\ref{tab:fit}.
    \begin{table}[tb]
      \centering
      \caption{Summary of the mean photon counts and resolutions for peaks in Fig.~\ref{fig:photon_spectrum}.}
      \begin{tabular}{c|c|c|c}
       & Energy & photon counts & resolution [FWHM] \\
      \hline
      xenon $\mathrm{K_\alpha}$ & \SI{29.68}{\keV} & \num{1.54964+-0.00008e4} & \SI{4.431+-0.097}{\%} \\
      xenon $\mathrm{K_\beta}$ & \SI{33.62}{\keV} & \num{1.76133+-0.00019e4} & \SI{4.364+-0.023}{\%} \\
      $^{228}\mathrm{Ac}$ & \SI{911.2}{\keV} & \num{4.80452+-0.00031e5} & \SI{1.103+-0.017}{\%} \\
      environmental $^{40}\mathrm{K}$ & \SI{1461}{\keV} & \num{7.70436+-0.00182e5} & \SI{1.065+-0.066}{\%} \\
      Double escape of $^{208}\mathrm{Tl}$ \SI{2615}{\keV} & \SI{1593}{\keV} & \num{8.40014+-0.00034e5} & \SI{0.980+-0.009}{\%} \\
      $^{208}\mathrm{Tl}$ & \SI{2615}{\keV} & \num{1.37868+-0.00059e6} & \SI{0.672+-0.083}{\%} \\
      \end{tabular}
      \label{tab:fit}
    \end{table}
    By interpolating these results, the energy resolution at the $^{136}$Xe $0\nu\beta\beta$ Q value of \SI{2458}{\keV} was estimated.
    Two types of energy dependence were considered: a case in which statistical fluctuation dominate ($a\sqrt{E}$) and a case in which systematic errors proportional to the energy exist ($a\sqrt{E+bE^2}$).
    Figure~\ref{fig:resolution_fit} shows the result of the interpolation to the $0\nu\beta\beta$ Q value.
    The estimated energy resolution at the $0\nu\beta\beta$ Q value is \SI{0.678+-0.010}{\%} for $a\sqrt{E}$ and \SI{0.788+-0.077}{\%} for $a\sqrt{E+bE^2}$.

    \subsection{Track topology}
\begin{figure}[tb]
      \centering
      \includegraphics[width=0.95\linewidth]{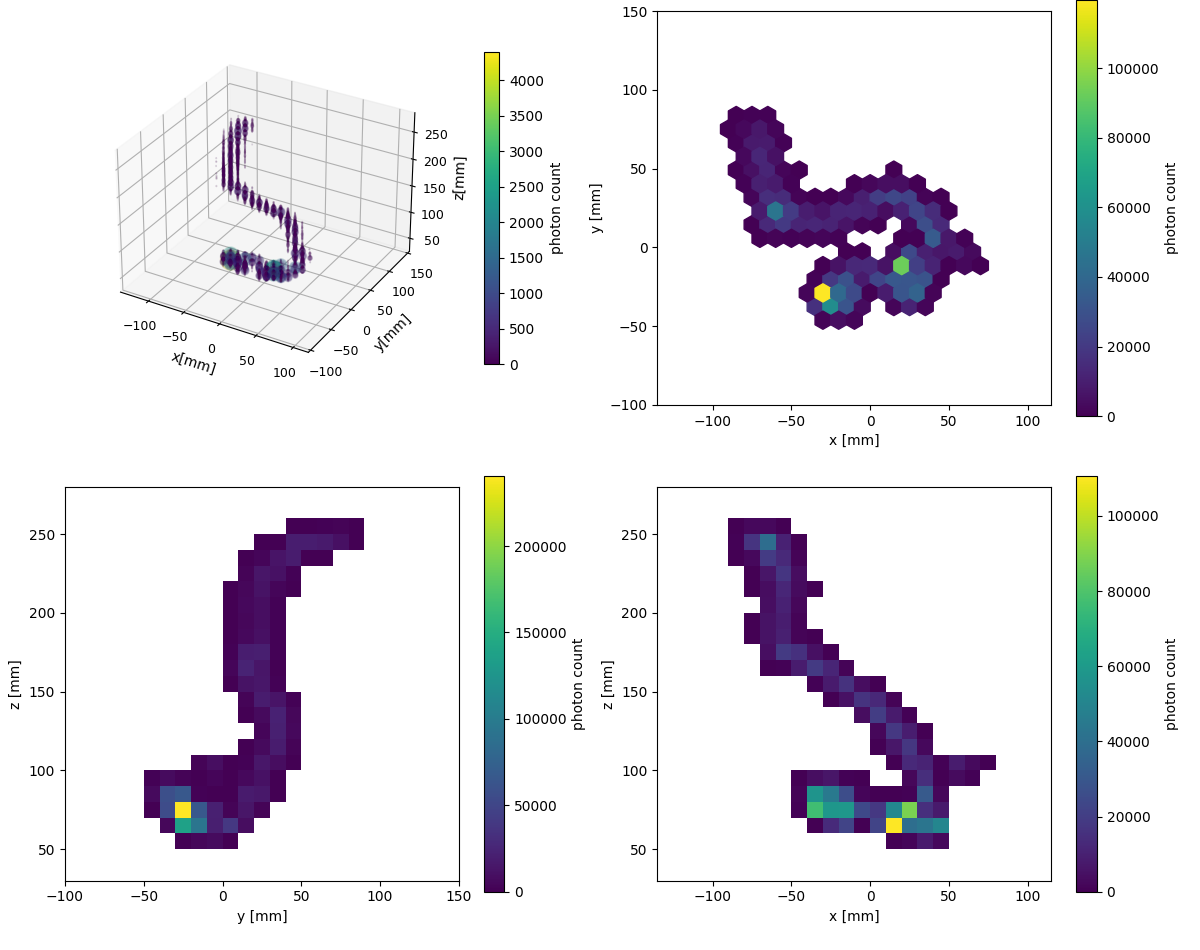}
      \caption{A sample track of \SI{2615}{\keV} event. In the 3D plot, the size of the points is proportional to the photon counts. One blob can be seen at the end of the track.}
      \label{fig:track_2615keV}
    \end{figure}

\begin{figure}[tb]
      \centering
      \includegraphics[width=0.95\linewidth]{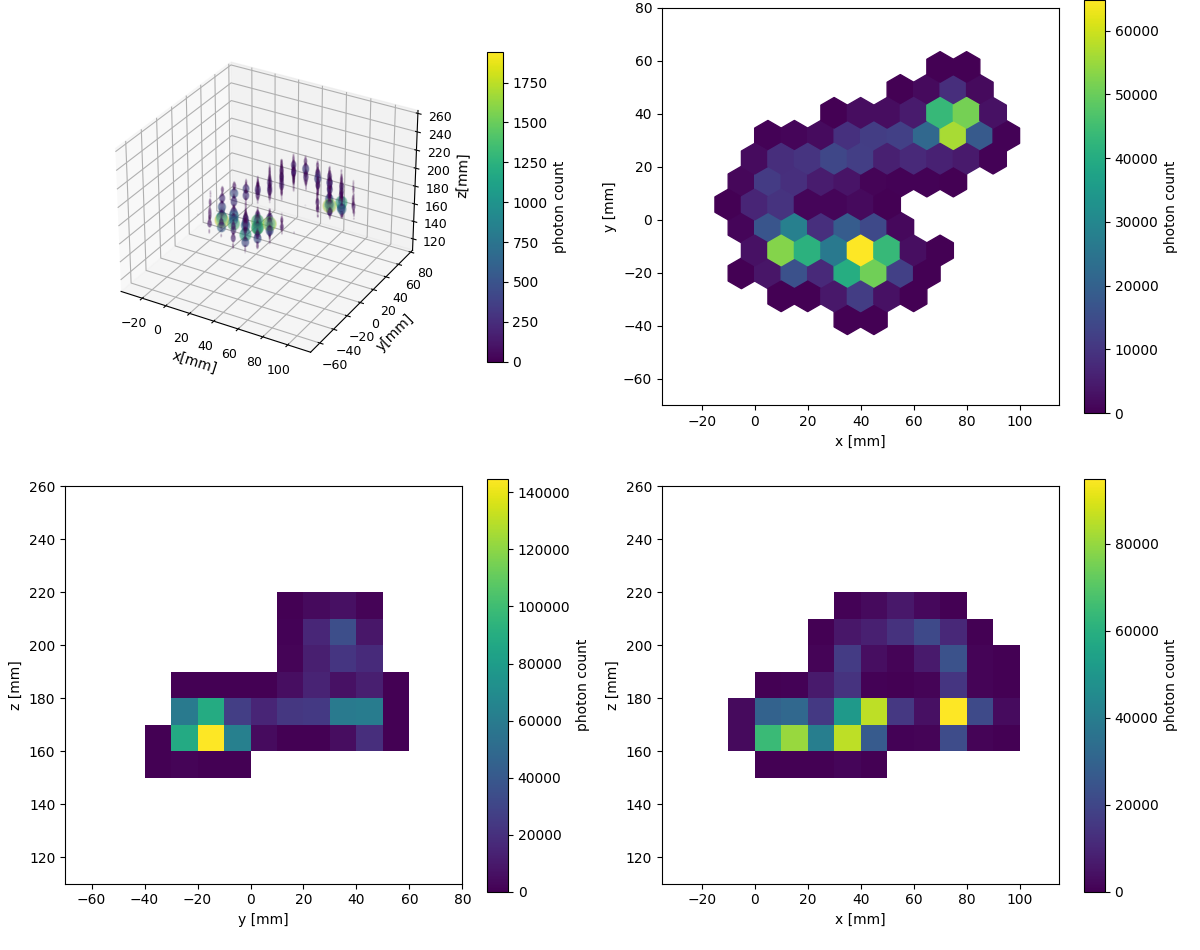}
      \caption{A sample track of \SI{1593}{\keV} event. In the 3D plot, the size of the points is proportional to the photon counts. There are 2 blobs corresponds to the end of the track of an electron and a positron.}
      \label{fig:track_1593keV}
    \end{figure}

    Reconstructions of a \SI{2615}{\keV} and a \SI{1593}{\keV} gamma induced track are shown in Fig.~\ref{fig:track_2615keV} and Fig.~\ref{fig:track_1593keV}, respectively.
    The \SI{2615}{\keV} event is considered to be due to the photoelectric absorption of gamma rays from $^{208}$Tl, and it is thought to be a single electron track.
    The \SI{1593}{\keV} event is considered to be due to the double escape of the \SI{2615}{\keV} gamma rays from $^{208}$Tl, and it is thought to be two tracks, one electron and one positron, originating from a single vertex due to pair creation.
    These events are selected from the photoelectric absorption peak and the double escape peak of \SI{2615}{\keV} gamma rays of $^{208}$Tl respectively.
    At the end of the track, there is a ``blob'' associated with high energy loss.
    In the \SI{2615}{\keV} event, there is one blob produced at the electron stop.
    The \SI{1593}{\keV} event produces two blobs at the end of electron and positron track, respectively.
    In events consisting of only one cluster, most events of \SI{2615}{\keV} appeared as single electron track as shown in Fig.~\ref{fig:track_2615keV}, while at \SI{1593}{\keV}, they appeared as tracks with one electron and one positron as shown in Fig.~\ref{fig:track_1593keV}.
    Since two blobs occur in the $0\nu\beta\beta$ event as well, we expect that this track information can be used to distinguish the single-electron events from gamma ray backgrounds when the blobs are cleary seen as in these figures.

\section{Conclusion}
We have newly developed a CW multiplier to generate high voltage for the AXEL detector, and installed it into the \SI{180}{\L} prototype detector.

To reduce its size the CW multiplier was made with surface-mount components on a flexible printed circuit board.
The outgassing rate of the \SI{180}{\L} prototype detector using the CW multiplier was \SI{1.5e-4}{\Pa.\m^3/\s}, indicating that gas purification could be performed without problems.
Surface discharge on the circuit was suppressed by applying additional insulating coating.

The 40 stage CW multiplier was installed in the \SI{180}{\L} detector.
Over 40 days of data taking with \SI{34.3}{\kV}, an energy resolution of \mbox{\SI{0.67+-0.08}{\%}} at \SI{2615}{\keV} was achieved and 3D tracks were reconstructed.
The AC pickup from the CW multiplier superimposed on the ELCC waveform is within one ADC count, confirming that its impact on the energy resolution is sufficiently small.
We have demonstrated 40 days of stable operation of the CW multiplier in the high-pressure xenon gas TPC.
  
\section*{Acknowledgements}
This work was supported by the JSPS KAKENHI Grant Numbers
18H05540, 
18J00365, 
18J20453, 
19K14738, 
20H00159, 
20H05251, 
and the JST SPRING, Grant Number
JPMJSP2114. 
We also appreciate the support for our project by Institute for Cosmic Ray Research, the University of Tokyo.
The development of the front-end board AxFEB is supported by Open-It (Open Source Consortium of Instrumentation).
\bibliographystyle{myptephy}
\bibliography{bibfile}

\end{document}